\documentclass[smallcondensed]{svjour3}
\usepackage{booktabs}
\usepackage{url}
\usepackage{amssymb}
\usepackage{amsmath}
\usepackage{algorithm,algorithmic}
\usepackage{wrapfig}
\usepackage{enumitem}
\usepackage{mdwlist}
\usepackage[dvipsnames]{xcolor}
\usepackage[caption=false]{subfig}
\usepackage{microtype}
\usepackage{color}
\usepackage{capt-of}
\usepackage{makecell}
\usepackage{float}
\usepackage[numbers]{natbib}
\usepackage{algorithm}
\usepackage{forest}
\usepackage{tikz}
\usepackage{rotating}
\usepackage{shellesc}
\usepackage{tablefootnote}
\usepackage{makecell}
\usetikzlibrary{shadows,positioning,arrows,automata,backgrounds,decorations,fit,petri,positioning,petri,shapes,calc,patterns,graphs}
\newlength{\hatchspread}
\newlength{\hatchthickness}
\newlength{\hatchshift}
\newcommand{\hatchcolor}{}
\tikzset{hatchspread/.code={\setlength{\hatchspread}{#1}},
	hatchthickness/.code={\setlength{\hatchthickness}{#1}},
	hatchshift/.code={\setlength{\hatchshift}{#1}},
	hatchcolor/.code={\renewcommand{\hatchcolor}{#1}}}
\tikzset{hatchspread=3pt,
	hatchthickness=0.4pt,
	hatchshift=0pt,
	hatchcolor=black}
\pgfdeclarepatternformonly[\hatchspread,\hatchthickness,\hatchshift,\hatchcolor]
{custom north west lines}
{\pgfqpoint{\dimexpr-2\hatchthickness}{\dimexpr-2\hatchthickness}}
{\pgfqpoint{\dimexpr\hatchspread+2\hatchthickness}{\dimexpr\hatchspread+2\hatchthickness}}
{\pgfqpoint{\dimexpr\hatchspread}{\dimexpr\hatchspread}}
{
	\pgfsetlinewidth{\hatchthickness}
	\pgfpathmoveto{\pgfqpoint{0pt}{\dimexpr\hatchspread+\hatchshift}}
	\pgfpathlineto{\pgfqpoint{\dimexpr\hatchspread+0.15pt+\hatchshift}{-0.15pt}}
	\ifdim \hatchshift > 0pt
	\pgfpathmoveto{\pgfqpoint{0pt}{\hatchshift}}
	\pgfpathlineto{\pgfqpoint{\dimexpr0.15pt+\hatchshift}{-0.15pt}}
	\fi
	\pgfsetstrokecolor{\hatchcolor}
	\pgfusepath{stroke}
}
\tikzset{
	petrinet/.style={font=\scriptsize,>=latex,			
		every edge/.append style={rounded corners},
		every token/.append style={minimum size=1mm},
		place/.style={circle,draw=black,inner sep=0pt,minimum size=4mm},
		finalplace/.style={place,pattern=custom north west lines,hatchspread=1.5pt,hatchthickness=0.25pt,hatchcolor=gray},
		transition/.style={rectangle,draw=black,minimum size=5mm},
		transitionlong/.style={rectangle,draw=black,minimum size=5mm},
		tau/.style={transition,fill=gray,inner sep=0pt},
		tau-transition-vertical/.style={tau,minimum height=5mm,minimum width=1mm},
		tau-transition-horizontal/.style={tau,minimum height=1mm,minimum width=5mm},
		tau-transition/.style={tau-transition-vertical},
		caption/.style={align=center, outer sep=0pt},		
	}
}
\tikzset{small/.style={level distance=10pt,sibling distance=-2pt,outer ysep=-2pt, label distance=-8pt}}
\tikzset{smaller/.style={level distance=20pt}}
\tikzset{smallish/.style={level distance=20pt,sibling distance=-2pt,outer ysep=-2pt, label distance=-8pt}}
\tikzset{node distance=5mm} 
\tikzset{place/.append style={circle,draw=black,thick,inner sep=0pt,minimum size=4mm,label position=below}}
\tikzset{transition/.append style={rectangle,draw=black,thick,inner sep=0pt,minimum size=6.8mm}}
\tikzset{tau/.style={transition,fill=black}}  
\definecolor{PuOr1}{RGB}{241,163,64}
\definecolor{PuOr2}{RGB}{247,247,247}
\definecolor{PuOr3}{RGB}{153,142,195}

\newcommand{\fleche}{\longrightarrow}
\newcommand{\flsup}[1]{\stackrel{#1}{\fleche}}
\newcommand{\step}[1]{\flsup{#1}}           
\newcommand{\Lan}                 {\mathfrak{L}}

\newcommand{\APN}{\mathit{APN}}

\newcolumntype{C}[1]{>{\centering\let\newline\\\arraybackslash\hspace{0pt}}m{#1}}

\smartqed  
\usepackage{graphicx}

\begin{document}

\title{Mining Non-Redundant Local Process Models From Sequence Databases}


\author{Niek Tax         \and
        Marlon Dumas 
}


\institute{N. Tax \at
              Eindhoven University of Technology, The Netherlands \\
              \email{n.tax@tue.nl}           
           \and
           M. Dumas \at
              University of Tartu, Estonia\\
              \email{marlon.dumas@ut.ee}
}

\date{Received: date / Accepted: date}

\maketitle

\begin{abstract}
Sequential pattern mining techniques extract patterns corresponding to frequent subsequences from a sequence database. A practical limitation of these techniques is that they overload the user with too many patterns. Local Process Model (LPM) mining is an alternative approach coming from the field of process mining. While in traditional sequential pattern mining, a pattern describes one subsequence, an LPM captures a set of subsequences. Also, while traditional sequential patterns only match subsequences that are observed in the sequence database, an LPM may capture subsequences that are not explicitly observed, but that are related to observed subsequences. In other words, LPMs generalize the behavior observed in the sequence database.
These properties make it possible for a set of LPMs to cover the behavior of a much larger set of sequential patterns. Yet, existing LPM mining techniques still suffer from the pattern explosion problem because they produce sets of redundant LPMs. In this paper, we propose several heuristics to mine a set of non-redundant LPMs either from a set of redundant LPMs or from a set of sequential patterns. We empirically compare the proposed heuristics between them and against existing (local) process mining techniques in terms of coverage, redundancy, and complexity of the produced sets of LPMs.

\keywords{Sequential Pattern Mining \and Process Mining \and Process Model}
\end{abstract}

\section{Introduction}
\label{sec:introduction}
Collections of sequences, also known as \emph{sequence databases}, are a common data source for knowledge extraction in many domains. Consider for example DNA and protein sequences, business process execution traces, customer purchasing histories, and software execution traces. Accordingly, the task of mining frequent patterns from sequence databases, known as \emph{sequential pattern mining}, is a mainstream research area in data mining. Originally, sequential pattern mining focused on extracting patterns that capture consecutive subsequences that recur in many sequences of a sequence database~\citep{Agrawal1995,Han2001,Zaki2001}. More recently, algorithms have been proposed to mine \emph{gapped} sequential patterns \citep{Ding2009,Tong2009,Wu2014}, which allow gaps between two successive events of the pattern, and \emph{repetitive} sequential patterns \citep{Ding2009,Tong2009}, which capture not only subsequences that recur in multiple sequences, but also subsequences that recur frequently within the same sequence.\looseness=-1

While sequential patterns may generate interesting insights, their practical application for data exploration is hindered by the fact that they often overload the analyst with a large number of patterns. To address this issue, several approaches have been proposed to describe sequence databases with smaller sets of patterns. One approach is to mine \emph{closed} patterns \citep{Ding2009,Yan2003} -- patterns for which there does not exist an extension with the same support. Another approach is to mine \emph{maximal} patterns \citep{Fournier2014,Zhang2001} -- patterns for which there does not exist an extension that meets the support threshold. Yet another approach is to mine \emph{compressed} patterns following the \emph{minimal description length} principle~\citep{Lam2014,Tatti2012}.
These approaches however are limited by the fact that each pattern captures only one frequent subsequence.

In alternative approaches, a pattern may capture multiple subsequences, including subsequences that are not observed in the sequence database, but that are related to observed subsequences. In other words, the extracted patterns generalize the observed behavior. For example, \emph{episodes} \citep{Mannila1997} extend sequential patterns with parallelism by allowing a pattern to incorporate partial order relations.
A recently proposed type of pattern that goes beyond the generalization capabilities of episodes (and other related work) is the \emph{Local Process Model (LPM)} \citep{Tax2016b}. An LPM is a pattern consisting of an arbitrary combination of sequence, parallelism, choice, and loop constructs. LPMs are represented as \emph{process trees} \citep{Buijs2012} -- a tree-based process modeling notation -- or as Petri nets~\citep{Murata1989}. \citet{Tax2016b} proposes a method to mine LPMs by iteratively expanding smaller LPMs into larger \emph{candidate LPMs}, followed by a step to evaluate the generated candidate LPMs. The approach shares common traits with the CloGSgrow algorithm for sequential pattern mining~\citep{Ding2009}, in that it mines gapped patterns and uses a notion of \emph{repetitive support}, meaning that it counts multiple occurrences of a pattern within the same sequence. 

Given their properties, it is possible for a small set of LPMs to cover the behavior of a much larger set of sequential patterns. However, the original LPM mining technique \citep{Tax2016b} still suffers from the pattern explosion problem because it is designed to extract one LPM at a time (in isolation). When applied repeatedly to a sequence database, this algorithm leads to a set of redundant LPMs.\looseness=-1

\begin{table}[t]
	\centering
	\vspace{-0.1cm}
	\caption{\emph{(a)} An example sequence database, and \emph{(b)} the patterns extracted with CloGSgrow \citep{Ding2009} using $\mathit{min\_sup}=3$.}
	\vspace{-0.1cm}
	\label{tab:running_example}
	\resizebox{\linewidth}{!}{
		\subfloat[]{
			\begin{tabular}{l|l}
				\toprule
				ID & Sequence\\
				\midrule
				1&$\langle \overline{\textcolor{blue}{E,B,A,B,A,F}},\overline{\textcolor{red}{A,C,B,D}}\rangle$\\
				2&$\langle \overline{\textcolor{blue}{E,B,A,F}},\overline{\textcolor{blue}{E,B,A,B,A,F}}\rangle$\\
				3&$\langle \overline{\textcolor{red}{A,B,C,D}},\overline{\textcolor{red}{A,C,D,B}},\overline{\textcolor{blue}{E,F}}\rangle$\\
				4&$\langle \overline{\textcolor{red}{A,C,D,B}},E,\overline{\textcolor{blue}{E,B,A,F}}\rangle$\\
				\bottomrule
		\end{tabular}}
		\quad
		\subfloat[]{
			\begin{tabular}{l|l|l|l|l|l}
				\toprule
				Sup & Pattern & Sup & Pattern & Sup & Pattern\\
				\midrule
				10&$\langle A \rangle$&4&$\langle A,C,D \rangle$&3&$\langle B,B,A,F \rangle$\\
				6&$\langle A,A \rangle$&3&$\langle A,E,F \rangle$&4&$\langle B,B,A \rangle$\\
				7&$\langle A,B \rangle$&10&$\langle B \rangle$&4&$\langle B,B,F \rangle$\\
				5&$\langle A,B,A \rangle$&7&$\langle B,A \rangle$&3&$\langle B,E,F \rangle$\\
				3&$\langle A,B,A,B \rangle$&5&$\langle B,A,B \rangle$&4&$\langle D \rangle$\\
				4&$\langle A,B,A,F \rangle$&3&$\langle B,A,B,F \rangle$&6&$\langle E \rangle$\\
				4&$\langle A,B,B \rangle$&5&$\langle B,A,F \rangle$&3&$\langle E,B,A,B,A \rangle$\\
				3&$\langle A,B,B,F \rangle$&4&$\langle B,A,A\rangle$&4&$\langle E,B,A,F \rangle$\\
				3&$\langle A,C,B \rangle$&6&$\langle B,B \rangle$&5&$\langle E,F \rangle$\\
				\bottomrule
		\end{tabular}}
	}
	\vspace{-0.5cm}
\end{table}
\begin{figure}[t]
	\centering
	\includegraphics[width=0.87\linewidth]{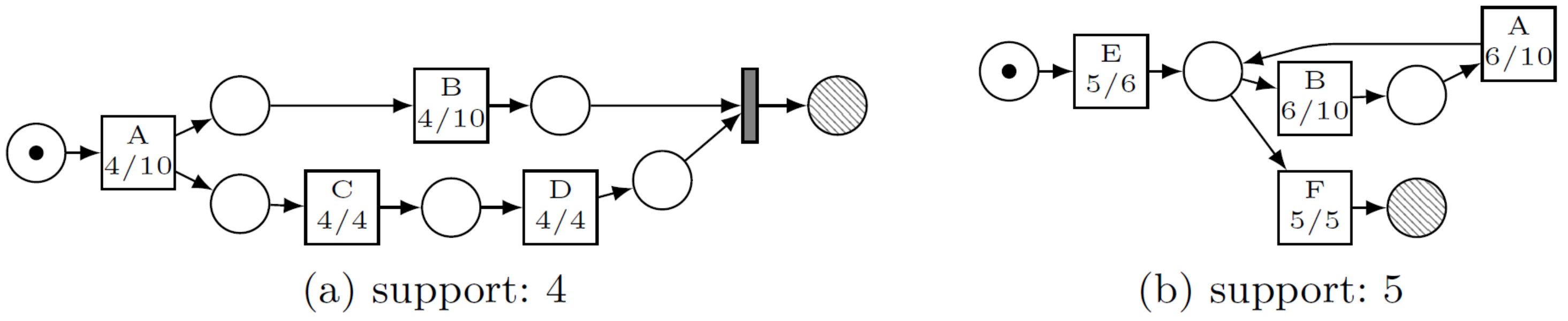}\\
	\vspace{-0.1cm}
	\caption{Two of the 717 Local Process Models (LPMs) mined from the sequence database of Table~\ref{tab:running_example}a using $\mathit{min\_sup}=3$ (visualized as Petri nets).}
	\label{fig:example_lpms}
	\vspace{-0.4cm}
\end{figure}

To illustrate the limitations of existing LPM mining and iterative sequential pattern mining techniques, we consider the sequence database shown in 
Table~\ref{tab:running_example}a. Table~\ref{tab:running_example}b shows the nine patterns produced by the CloGSgrow algorithm \citep{Ding2009} with a minimum support of three. In total, CloGSgrow requires 29 patterns to describe the behavior in this sequence database. Applying basic LPM mining with a minimum support of three leads to 717 patterns, two of which are shown in Figure~\ref{fig:example_lpms}. The LPM of Figure~\ref{fig:example_lpms}a (LPM~\emph{(a)}) expresses that $A$ is followed by $B$, $C$, and $D$, where the $D$ can only occur after $C$, and the $B$ can occur at any point after $A$. The LPM of Figure~\ref{fig:example_lpms}b (LPM~\emph{(b)}) is equivalent to regular expression E(BA)*F. 
The numbers printed in the LPMs respectively indicate the number of events explained by the LPM patterns and the number of occurrences of each activity in the sequence database, e.g., 4 out of 10 occurrences of activity $A$ in the sequence database are explained by LPM~\emph{(a)}. The four instances of LPM~\emph{(a)} are indicated in \textcolor{red}{red} in Table~\ref{tab:running_example}a, and the five instances of LPM~\emph{(b)} are indicated in \textcolor{blue}{blue}. LPMs~\emph{(a)} and \emph{(b)} together describe almost all behavior in the sequence database in a compact manner. While basic LPM mining with a minimum support of three results in 717 patterns, the desired output would be only the two LPMs of Figure 1.

In this setting, the contributions of this article are:
\begin{itemize}
	\item A framework to evaluate the coverage, redundancy, and complexity of a set of Local Process Models with respect to a sequence database. 
	\item	Alternative heuristics to mine a set of non-redundant LPMs from a sequence database, by post-processing either a set of redundant LPMs or a set of gapped sequential patterns such as those produced by CloGSgrow.
\end{itemize}

One application that we envision for mining non-redundant LPMs is the identification of repetitive routines that may be amenable for automation using Robotic Process Automation (RPA) tools~\cite{Aalst2018}. 
These tools allow us to record and replay routines consisting of low-level actions such as opening an application, copying and pasting data from/into fields in a Web form, copying and pasting data from/into spreadsheet applications, etc.  For example, an RPA tool allows us to automate the set of actions performed by a clerk when handling a purchase order or the actions performed by a Human Resources officer when a new employee starts to work.
A major practical question in the context of RPA projects is how to identify candidate routines for automation.
To answer this question, analysts currently rely on expert judgment and manual observation of process workers. 
An alternative and more systematic approach is to record the actions that process workers perform during a work shift in the form of \emph{event logs} (i.e. databases of event sequences) and to extract patterns from such logs.
The resulting set of patterns must be small (in order not to overwhelm the analyst), each pattern must occur frequently, and each pattern must generalize the behavior recorded in the event log, since the event log most likely does not contain all possible executions of an activity. Achieving these desirable properties is the main aim of the LPM mining heuristics introduced in this article. Another potential application where these properties are desirable is that of mining daily habits and routines from smart home data, such that the smart home system can be configured to support these daily habits and routines~\citep{Tax2018b}. Accordingly, we empirically evaluate the proposed heuristics for mining non-redundant LPMs using real-life sequence databases covering both business process execution logs and smart home sensor logs. 


This paper is structured as follows. Section~\ref{sec:background} introduces background concepts related to process models, LPMs, and automated process discovery, and discusses related work. Section~\ref{sec:quality_criteria} outlines quality criteria for LPM sets. Section~\ref{sec:mining_approaches} presents the proposed heuristics to discover sets of non-redundant LPMs. Section~\ref{sec:experiments} presents an empirical evaluation of the proposed heuristics using real-life sequence databases. Finally, Section~\ref{sec:conclusions} draws conclusions and outlines directions for future work.

\section{Background}
\label{sec:background}
In this section, we introduce notation and basic concepts related to sequence databases, process models, process discovery, and Local Process Models (LPMs).
\vspace{-0.3cm}
\subsection{Events, Sequences, and Sequence Databases}
Let $X^*$ denote the set of all sequences over a set $X$ and $\sigma{=}\langle a_1,a_2,\dots,a_n\rangle$ a sequence of length $n$, with $\sigma(i){=}a_i$ and $|\sigma|{=}n$. $\langle\rangle$ is the empty sequence and $\sigma_1{\cdot}\sigma_2$ is the concatenation of sequences $\sigma_1$ and $\sigma_2$. We denote with $\sigma{\upharpoonright}_{X}$ the projection of sequence $\sigma$ on set $X$, e.g., for $\sigma{=}\langle a,b,c\rangle$, and $X{=}\{a,c\}$, $\sigma{\upharpoonright}_{X}{=}\langle a,c\rangle$. Likewise, $\sigma{\downharpoonleft}_{X}$ indicates sequence $\sigma$ where all members of $X$ are filtered out, e.g., $\langle a,b,c,b\rangle{\downharpoonleft}_{\{b\}}{=}\langle a,c\rangle$. $\mathit{hd}^k(\sigma){=}\langle a_1, a_2, \dots, a_k\rangle$ is the prefix of length $k$ (with $0 {<} k {<} |\sigma|$) of sequence $\sigma$, for example, $\mathit{hd}^2(\langle a,b,c,d,e\rangle){=}\langle a,b\rangle$. A multiset (or bag) over $X$ is a function $B:X{\rightarrow}\mathbb{N}$ which we write as $[a_1^{w_1},a_2^{w_2},\dots,a_n^{w_n}]$, where for $1{\le} i {\le} n$ we have $a_i{\in} X$ and $w_i{\in}\mathbb{N}^{+}$. The set of all multisets over $X$ is denoted $\mathcal{B}(X)$.

An \emph{event} $e$ (also called \emph{symbol}) denotes the occurrence of an activity. We write $\Sigma$ to denote the set of all possible activities (also called the \emph{alphabet of symbols}). An event sequence (called a \emph{trace} in the process mining field) is a sequence $\sigma{\in}\Sigma^*$. A \emph{sequence database} (called an \emph{event log} in the process mining field) is a finite multiset of sequences, $\mathit{SD}{\in}\mathcal{B}({\Sigma^*})$. For example, the sequence database $\mathit{SD}{=}[\langle a,b,c\rangle^2,\langle b,a,c\rangle^3]$ consists of two occurrences of sequences $\langle a,b,c\rangle$ and three occurrences of sequence $\langle b,a,c\rangle$. Finally, we lift the operations for projection and filtering of sequences to multisets of sequences, i.e., $\mathit{SD}{\upharpoonright}_{\{a,c\}}{=}\mathit{SD}{\downharpoonleft}_{\{b\}}{=}[\langle a,c\rangle^5]$. 

\subsection{Process Models}
We use Petri nets to represent process models due to their formal semantics. A Petri net is a directed bipartite graph consisting of places (depicted as circles) and transitions (depicted as rectangles), connected by arcs. A transition describes an activity, while places represent the enabling conditions of transitions. Labels of transitions indicate the type of activity that they represent. Unlabeled transitions ($\tau$-transitions) represent invisible transitions (depicted as gray rectangles), which are only used for routing purposes and are not recorded in the sequence database.\looseness=-1
\begin{definition}[Labeled Petri net]
	\label{def:lpn}
	A \emph{labeled Petri net} $N=\langle P,T,F,\ell\rangle$ is a tuple where $P$ is a finite set of places, $T$ is a finite set of transitions such that $P{\cap}T{=}\emptyset$,  $F{\subseteq}(P {\times}T){\cup}(T{\times}P)$ is a set of directed arcs, called the flow relation, and $\ell{:}T{\nrightarrow}\Sigma$ is a partial labeling function that assigns a label to a transition, or leaves it unlabeled (the $\tau$-transitions). We write $\bullet{n}$ and $n\bullet$ for the input and output nodes of $n\in P \cup T$ (according to $F$).
\end{definition}

An example of a labeled Petri net is shown in Figure~\ref{fig:example_lpms}a, consisting of five places and five transitions out of which four are labeled transitions (labeled $A$, $B$, $C$, and $D$) and one is a $\tau$-transition.

The state of a Petri net is defined by its \emph{marking} $m{\in} \mathcal{B}(P)$ being a multiset of places. A marking is graphically denoted by putting $m(p)$ tokens on each place $p{\in}P$, e.g., the Petri net in Figure~\ref{fig:example_lpms}a contains one token in the leftmost place. State changes occur through transition firings. A transition $t$ is enabled (can fire) in a given marking $m$ if each input place $p{\in}{\bullet}t$ contains at least one token. Once $t$ fires, one token is removed from each input place $p{\in}{\bullet} t$ and one token is added to each output place $p'{\in}t \bullet$, leading to a new marking $m'{=}m{-}\bullet{t}+t\bullet$. In the Petri net of Figure~\ref{fig:example_lpms}a, the transition that is labeled $A$ is enabled from the indicated marking, and firing this transition leads to a marking from which two transitions are enabled, which are labeled $B$ and $C$. A firing of a transition $t$ leading from marking $m$ to marking $m'$ is denoted as step $m {\step{t}} m'$. Steps are lifted to sequences of firing  enabled transitions, written $m {\step{\gamma}} m'$ and $\gamma {\in}T^*$ is a \emph{firing sequence}.\looseness=-1

Defining an \emph{initial} and \emph{final} markings allows to define the \emph{language} that is accepted by a Petri net as a set of finite sequences of activities. To define this language of a Petri net, we need to lift the labeling partial function $\ell$ to sequences to be able to apply it to the firing sequences that are allowed by the Petri net. 
A partial function $f{\in} X {\nrightarrow} Y$ with domain $\mathit{dom}(f)$ can be lifted to sequences over $X$ using the following recursion: (1) $f(\langle\rangle){=}\langle\rangle$;  (2) for any $\sigma{\in} X^*$ and $x{\in}X$:
\begin{center}
	$f(\sigma \cdot \langle x\rangle) =
	\left\{
	\begin{array}{ll}
	f(\sigma)  & \mbox{if } x{\notin}\mathit{dom}(f), \\
	f(\sigma) \cdot \langle f(x)\rangle & \mbox{if } x{\in}\mathit{dom}(f).
	\end{array}
	\right.$
\end{center}

The combination of a labeled Petri net with an initial and final marking, for which we can define the language, we will refer to as an \emph{accepting Petri Net}. 

\begin{definition}[Accepting Petri Net]
	An \emph{accepting Petri net} is a triplet $\APN=(N,m_0,m_f)$, where $N$ is a labeled Petri net, $m_0{\in}\mathcal{B}(P)$ is its initial marking, and $m_f{\in}\mathcal{B}(P)$ its final marking. A sequence $\sigma{\in}\Sigma^*$ is a \emph{trace} of an accepting Petri net $\APN$ if there exists a firing sequence $m_0{\step{\gamma}}m_f$, $\gamma{\in}T^*$ and $\ell(\gamma){=}\sigma$.
\end{definition}
In this paper, places that belong to the initial marking contain a token and places belonging to the final marking are marked as $\begin{tikzpicture}
[node distance=1.4cm,
on grid,>=stealth',
bend angle=20,
auto,
every place/.style= {minimum size=0.0mm},
]
\node [place,pattern=custom north west lines,hatchspread=1.5pt,hatchthickness=0.25pt,hatchcolor=gray] {};
\end{tikzpicture}$. For example, the initial marking of the Petri net shown in Figure~\ref{fig:example_lpms}a consists of a single place (i.e., the leftmost place) and its final marking also consists of a single place (i.e., the rightmost place).

The \emph{language} $\Lan(\APN)$ is the set of all its traces, i.e., $\Lan(\APN)=\{l(\gamma) | \allowbreak \gamma{\in}T^*\land m_0{\step{\gamma}}m_f\}$, which can be of infinite size when $\APN$ contains loops. For example, the language of the accepting Petri net shown in Figure~\ref{fig:example_lpms}a is $\{\langle A,B,C,D\rangle,\langle A,C,B,D\rangle,\langle A,C,D,B\rangle\}$. While we have now defined the language of accepting Petri nets, in theory, $\Lan(M)$ can be defined for any process model $M$ regardless of its notation, as long as the notation has formal semantics. We denote the universe of process models as $\mathcal{M}$. For each $M{\in}\mathcal{M}$, $\Lan(M)\subseteq\Sigma^*$ is defined.

\subsection{Automated Process Discovery}
Automated Process Discovery methods \citep{Aalst2016} are concerned with the extraction of a (single) process model $M\in\mathcal{M}$ that describes the behavior observed in, or implied by, a sequence database $\mathit{SD}$. A process discovery method is a function $\mathit{PD}:\mathcal{B}({\Sigma^*})\rightarrow\mathcal{M}$ that produces a process model from a sequence database (called an \emph{event log} in the process mining terminology). One of the early works on process discovery is the $\alpha$-algorithm~\cite{Aalst2004}, which first infers ordering relations between pairs of activities and then aims to find a Petri net that adheres to these binary relations. Recent techniques in process discovery are the Inductive Miner~\cite{Leemans2013} and the Split Miner~\cite{Augusto2017,Augusto2019}. 

\begin{figure}
	\centering
	\includegraphics[width=0.85\textwidth]{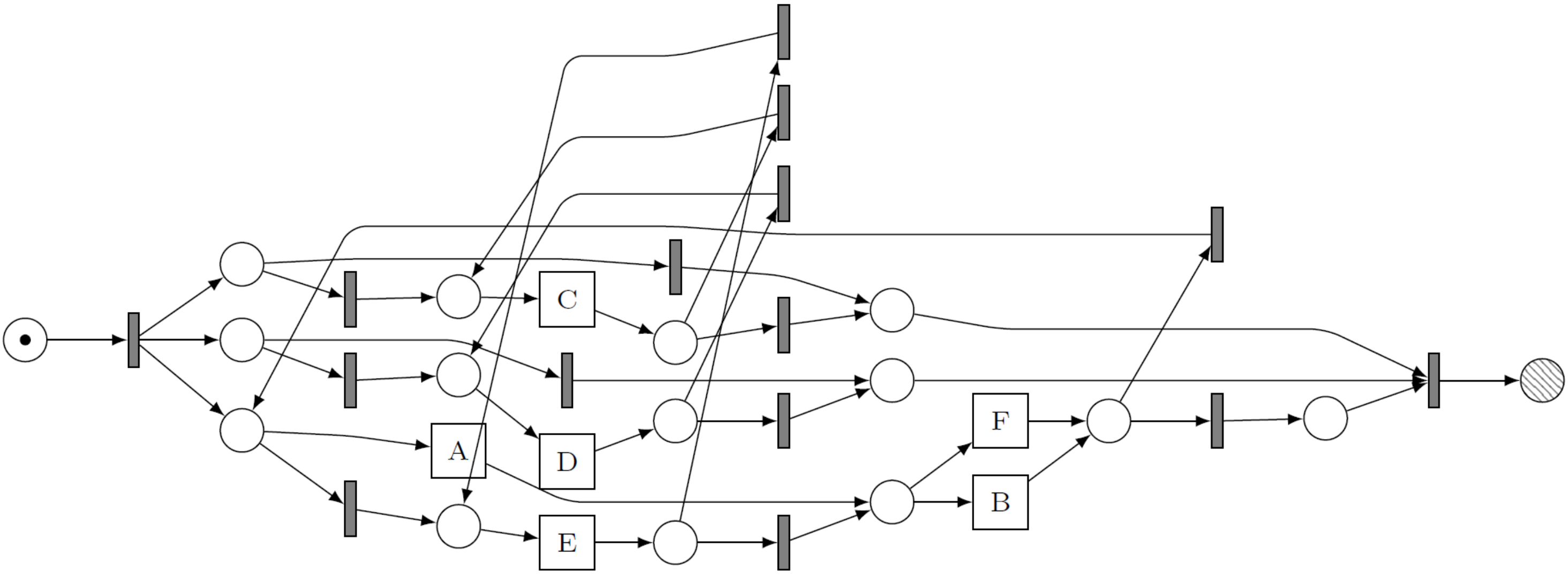}\\
	\caption{The resulting process model obtained by applying the Inductive Miner process discovery algorithm \citep{Leemans2013} to the sequence database of Table~\ref{tab:running_example}a.}
	\label{fig:im_result}
	\vspace{-0.1cm}
\end{figure}

The Inductive Miner \citep{Leemans2013} is a representative of this family of techniques. Figure~\ref{fig:im_result} shows the process model produced by the Inductive Miner when applied to the sequence database of Table~\ref{tab:running_example}a. While process discovery techniques produce useful results over simple sequence databases, they create process models that are either very complex or overgeneralizing (i.e., allowing for too much behavior when applied to real-life datasets
For example, the process model of Figure~\ref{fig:im_result} over-generalizes the sequence database in Table~\ref{tab:running_example}a. Indeed, the process model of Figure~\ref{fig:im_result} allows us to perform every combination of event A-F, with the only constraint that there has to be at least one occurrence of A or of E,  and each occurrence of A or B must be followed by an occurrence of F or B. Other symbols (C, D, and E) can occur any number of times and at any point in the sequence. 


Several measures have been developed to asses whether a discovered process model accurately describes the sequence database from which is has been discovered. First, the discovered process model should cover as much as possible of the behavior that was observed in the sequence database. \emph{Fitness} measures the degree to which the behavior in the sequence database is represented by the process model. The most widely used measure to quantity fitness is \emph{alignment-based fitness}~\cite{Aalst2012}. Additionally, the process model should not allow for too much behavior that was not observed in the sequence database. \emph{Precision} measures the amount of behavior that was modeled but that did in fact never happen. Several measures exist to measure precision, with \emph{escaping edges precision}~\cite{Munoz2010} being the most widely used one. To give a mathematically precise definition of fitness and precision we first define the \emph{trace set} of a sequence database. For a sequence database $\mathit{SD}$, $\tilde{\mathit{SD}}{=}\{\sigma{\in}\Sigma^*|\mathit{SD}(\sigma){>}0\}$ is the \emph{trace set} of $\mathit{SD}$, e.g., for sequence database $\mathit{SD}{=}[\langle a,b,c\rangle^2,\langle b,a,c\rangle^3]$, $\tilde{\mathit{SD}}{=}\{\langle a,b,c\rangle\langle b,a,c\rangle\}$.
For a sequence database $\mathit{SD}$ and a process model $M$, we say that $\mathit{SD}$ is \emph{fitting} on $M$ if $\tilde{\mathit{SD}}{\subseteq}\Lan(M)$. \emph{Precision} is related to the behavior that is allowed by a model $M$ that was not observed in the sequence database $\mathit{SD}$, i.e., $\Lan(M){\setminus}\tilde{\mathit{SD}}$.

\subsection{Local Process Models}
Local Process Models (LPMs) \citep{Tax2016b} are process models that describe frequent but partial behavior; i.e., they model a subset of the process activities that were seen in the sequence database. To allow for an apriori-like iterative expansion of patterns, LPMs are mined using a tree-based process modeling notation called \emph{process trees}. The iterative expansion procedure of LPM is often bounded to a maximum number of expansion steps (in practice often to 4 steps), as the number of possible expansions that need to be considered grows combinatorially with the number of activities in the sequence database as well as with the maximum LPM size. While LPMs are mined in the form of process trees, they can be easily converted and be represented in any process modeling notation, such as BPMN\footnote{\url{http://www.bpmn.org/}}, UML Actvity Diagrams\footnote{\url{http://www.omg.org/spec/UML/2.5/}}, or Petri nets \citep{Murata1989}. In this paper we use the latter due to their formal semantics.\looseness=-1

\begin{figure}[t]
	\centering
	\subfloat[]{
		\centering
		\begin{forest}
			[,name=n0 
			[A,name=n1]
			]
		\end{forest}
		\label{sfig:m0}
	} 
	\quad
	\subfloat[]{
		\centering
		\begin{forest}
			[$\rightarrow$,name=n0 	
			[A,name=n3]
			[B,name=n4]
			]
		\end{forest}
		\label{sfig:m1}
	} 
	\quad
	\subfloat[]{
		\centering
		\begin{forest}
			[$\rightarrow$,name=n0 
			[A,name=n1]
			[$\wedge$,name=n2
			[B,name=n3]
			[C,name=n4]
			]
			]
		\end{forest}
		\label{sfig:m2}
	}
	\quad
	\subfloat[]{
		\centering
		\begin{forest}
			[$\rightarrow$,name=n0 
			[$\times$,name=n4
			[A,name=n1]
			[D,name=n6]
			]
			[$\wedge$,name=n2
			[B,name=n3]
			[C,name=n5]	
			]
			]
		\end{forest}
		\label{sfig:m3}
	}
	\quad
	\subfloat[\label{sfig:example_alignment}]{
		\includegraphics[width=0.35\linewidth]{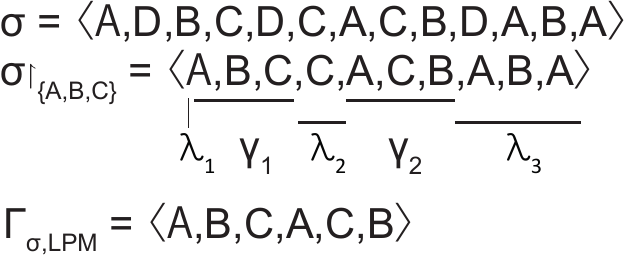}
	}
	\caption{\emph{(a)} An initial LPM $M_1$ and \emph{(b)} $M_2$, \emph{(c)} $M_3$, \emph{(d)} $M_4$, three LPM built from successive expansions. \emph{(e)} The segmentation of sequence $\langle A,D,B,C,D,C,A,C,B,D,A,B,A\rangle$ on $M_3$.}
	\label{fig:expansion}
	\vspace{-0.3cm}
\end{figure}

A process tree is a tree where leaf nodes represent activities, and non-leaf nodes represent \emph{operators} that specify the allowed behavior over the activity nodes. Supported operator nodes are the \textit{sequence} operator ($\rightarrow$) that indicates that the first child is executed before the second, the \textit{exclusive choice} operator ($\times$) that indicates that exactly one of the children can be executed, the \textit{concurrency} operator ($\wedge$) that indicates that every child will be executed without putting restrictions on the order, and the \emph{loop} operator ($\circlearrowright$), which has one child node and allows for repeated execution of this node. Like for other process modeling notations we can specify the language for process trees. Figure~\ref{sfig:m3} shows an example process tree $M_4=\rightarrow(\times(A,D),\wedge(B,C))$, where the $\rightarrow$ operator indicates that its left child $\times(A,D)$ is executed before its right child $\wedge(B,C)$. Left child $\times(A,D)$ contains a choice operator and therefore allows for either $A$ or $D$. Right child $\wedge(B,C)$ executes $B$ and $C$ in any order. Combined, $\Lan(\mathit{M_4}){=}\{\langle A,B,C\rangle,\langle A,C,B\rangle,\langle D,B,C\rangle, \langle D,C,B\rangle\}$, i.e., either activity A or D is executed first, followed by activities B and C in any order. Process trees can be trivially converted to Petri nets, e.g., process tree $\rightarrow(A,\wedge(B,\rightarrow(C,D)))$ transforms into the Petri net shown in Figure~\ref{fig:example_lpms}a.

An algorithm to generate a ranked list of LPMs via iterative expansion of candidate process trees is proposed in \citet{Tax2016b}. An expansion step of an LPM is performed by replacing one of the leaf nodes of the process tree by an operator node (i.e., $\rightarrow$,$\times$,$\wedge$, or $\circlearrowright$), where one of the child nodes is the activity of the replaced leaf node $a$ and the other is a new activity node $b\in\Sigma$. An LPM $M{\in}\mathcal{M}$ can be expanded in many ways, as it can be extended by replacing any one of its activity nodes, expanding it with any of the operator nodes, and with a new activity node that represents any of the activities in the sequence database. We define $Exp(M)$ as the set of expansions of $M$, and $\mathit{exp}\_\mathit{max}$ the maximum number of expansions allowed from an \emph{initial LPM}; i.e., an LPM containing only one activity.

To find the instances of a given LPM in a given sequence database $\mathit{SD}$ in order to count its \emph{support} and \emph{confidence}, its sequences $\sigma{\in}\mathit{SD}$ are first projected on the set of activities $X$ in the LPM, i.e., $\sigma'=\sigma{\upharpoonright}_{X}$. The projected sequence $\sigma'$ is then segmented into $\gamma$-segments that fit the behavior of the LPM and $\lambda$-segments that do not fit the behavior of the LPM, i.e., $\sigma'{=}\lambda_1\cdot\gamma_1\cdot\lambda_2\cdot\gamma_2\cdot\dotso\cdot\lambda_n\cdot\gamma_n\cdot\lambda_{n+1}$ such that $\gamma_i{\in}\Lan(\mathit{LPM})$ and $\lambda_i{\not\in}\Lan(\mathit{LPM})$. 
We define $\Gamma_{\mathit{LPM}}(\sigma)$ to be a function that projects sequence $\sigma$ on the LPM activities and obtains its subsequences that fit the LPM, i.e., $\Gamma_{\mathit{LPM}}(\sigma)=\gamma_1\cdot\gamma_2\cdot\dotso\cdot\gamma_n$.

Consider $M_3$ and sequence $\sigma$ from Figure \ref{fig:expansion}. Function $\mathit{Act}(\mathit{LPM})$ retrieves the set of process activities in the LPM, e.g., $\mathit{Act}(M_3){=}\{A,B,C\}$.
Projection on the activities of the LPM gives $\sigma{\upharpoonright}_{\mathit{Act}(M_3)}{=}\langle A,B,C,C,A,C,B,A,B,A\rangle$. 
Figure~\ref{sfig:example_alignment} shows the segmentation of the projected sequence on the LPM, leading to $\Gamma_{\mathit{LPM}}(\sigma)=\langle A,B,C,A,C,B\rangle$. 
The segmentation starts with an empty non-fitting segment $\lambda_1$, followed by a fitting segment $\gamma_1{=}\langle A,B,C\rangle$, which completes one run through the process tree. 
The second event $C$ in $\sigma$ cannot be replayed on $LPM$, since it only allows for one $C$ and $\gamma_1$ already contains a $C$. This results in a non-fitting segment $\lambda_2{=}\langle C\rangle$. $\gamma_2{=}\langle A,C,B\rangle$ again represents a run through process tree. The segmentation ends with non-fitting segment $\lambda_3{=}\langle A,B,A\rangle$. We lift segmentation function $\Gamma$ to sequence databases, $\Gamma_{\mathit{LPM}}(\mathit{SD}){=}\{\Gamma_{\mathit{LPM}}(\sigma)|\sigma{\in}\mathit{SD}\}$. An alignment-based \citep{Aalst2012} implementation of $\Gamma$, as well as a method to rank and select LPMs based on their support, i.e., the number of events in $\Gamma_{\mathit{LPM}}(\mathit{SD})$, is given in \citet{Tax2016b}. The time complexity of this implementation of $\Gamma$ is exponential in the number of activities as well as exponential in the length of the sequence.

\subsection{Related Work}
\label{sec:related_work}
The problem of mining sequential patterns that incorporate concurrency, choice, and repetition constructs has been an active field of study in the past two decades, starting with the work on \emph{episodes} \citep{Mannila1997}. Episodes extend sequential patterns with some form of concurrency by allowing a pattern to incorporate \emph{partial order relations}. In constast to sequential pattern mining, episode mining techniques traditionally addressed the task of mining patterns from a \emph{single sequence}. A recent approach addresses the mining of episodes from a sequence database (as opposed to a single sequence)~\citep{Leemans2014}. \citet{Harms2001} proposed a technique to mine closed episodes, while \citet{Pei2006} showed that episode mining techniques cannot mine arbitrary partial orders, and proposed a technique to mine closed \emph{partial order patterns}. 
In an alternative approach, \citet{Lu2009} proposed a method called Post Sequential Patterns Mining (PSPM) that takes as input a set of sequential patterns
and extracts concurrency relations between them. 
A later extension \citep{Lu2011} improves the procedure to extract concurrency relations and proposes a visual notation, called a ConSP-Graph, to represent the concurrency relations between sequential patterns. 
In ConSP-Graphs, concurrency relations between sequential patterns means that those patterns \emph{occur together in the same sequences}. This notion of concurrency is slightly different than the one supported in episodes, where concurrency relations exist between events, and not between patterns. In an LPM, a concurrency relation can exist both between multiple events (as in episodes) or between multiple sub-patterns as in ConSP-Graphs.

\citet{Diamantini2016a} describe a method to mine frequent patterns with concurrency via a two-step approach. First, the sequence database is transformed into a set of so-called \emph{instance graphs}, i.e.\ partially ordered set of symbols. In the second step, a graph clustering technique is applied to obtain frequent subgraphs from the set of instance graphs. Like episodes and partial order patterns, instance graphs can capture sequential and concurrency relations, but they cannot capture choice and loop constructs, as LPMs do.

\citet{Chen2010} extended the work by \citet{Lu2009} by extracting \emph{exclusive relations} between sequential patterns, i.e., patterns that do not occur in the same sequences, and proposed a visual graph called an ESP-graph to visually represent such relations. However, ESP-graphs cannot capture arbitrary combinations of concurrency, choices, sequential orderings, and loops.


Declarative process model discovery \citep{Ciccio2016,Maggi2012} is a subfield of automated process discovery, where the discovered process models consist of sets of Linear Temporal Logic (LTL) constraints over the activities in the process. 
Each constraint can be seen as a pattern that relates multiple activities via sequential, choice, and repetition relations. One limitation of this approach is that constraints in declarative process models do not include concurrency relations, since such relations cannot be captured using LTL. Also, to the best of our knowledge, none of the existing research on declarative process model discovery has addressed the question of discovering non-redundant or maximal constraints (i.e.\ patterns).


\citet{Saetrom2003} use genetic programming to mine patterns expressed in the regular-expression-like Interagon Query Language (IQL), thereby allowing the patterns to generalize to a higher degree than episodes and partial order patterns by additionally allowing patterns to incorporate choice and Kleene star (i.e., repetition) constructs. However, in IQL, concurrency relations can only be defined at the level of symbols (as in episodes) and not at the level of (sub-)patterns. Hence, IQL cannot capture a pattern where two symbols (say A and B) can be repeated any number of times, and in the middle of the repeated occurrences of A and B, a third symbol C may occur. The latter can be captured by an LPM with two concurrent branches: one branch consisting of a loop containing A followed by B, and the other branch  consisting of C, i.e. $\wedge(\circlearrowright(\rightarrow(A,B)),C)$. Another limitation of the methods in \citet{Saetrom2003} is that the discovered patterns may be redundant. In other words, the LPM mining methods proposed in this article extend IQL and other previous work by allowing us to extract more expressive patterns and ensuring that the resulting patterns are non-overlapping.


The original LPM mining algorithm~\cite{Tax2016b} discovers all LPMs with a support above a given threshold. In~\cite{Tax2016c,Tax2017}, we proposed approximate algorithms for LPM mining, which achieve higher computational efficiency by sacrificing the guarantee that all LPMs with a minimal support threshold are found. In more recent work, we proposed algorithms for mining a set of LPMs from a sequence database based on utility functions~\cite{Tax2018,Dalmas2018}. These approaches allow us to mine a set of LPMs that optimize a user-specified utility function and that additionally specify some user-specified constraints, e.g. mining the LPMs with the longest execution times, instead of mining the most frequently occurring LPMs. Later work, we presented efficient algorithms for mining LPMs under two specific types of constraint, called \emph{event gap constraints} and \emph{time gap constraints}~\cite{Tax2018b}. 
None of these previous studies however has addressed the problem of discovering non-redundant LPMs.


\section{Quality Criteria for Local Process Model Sets}
\label{sec:quality_criteria}
\begin{figure}[t]
	\centering
	\includegraphics[width=\linewidth]{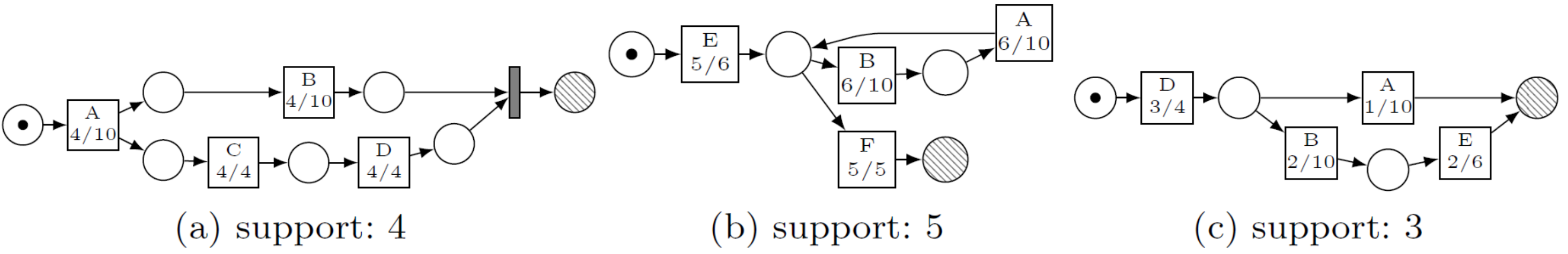}
	\caption{A Local Process Model set $\mathit{LPMS}$ consisting of three example Local Process Models mined from sequence database $\mathit{SD}$.} 
	\label{fig:local_process_models}
	\vspace{-0.3cm}
\end{figure}

We illustrate the need for quality criteria for Local Process Model (LPM) sets that takes into account the \emph{redundancy} of the pattern set through an example. Consider the sequence database $\mathit{SD}$ shown in Table 1a and and Local Process Model set $\mathit{LPMS}$ that consist of the three LPMs of Figure~\ref{fig:local_process_models}, to which, compared to Figure \ref{fig:example_lpms}, LPM \emph{(c)} is added as one example of a redundant LPM from the set of 717 LPMs. There is overlap in the activities that are described by the LPMs, e.g., LPM \emph{(a)}, \emph{(b)}, and \emph{(c)} all contain a transition that is labeled $A$, which means that there are multiple candidate patterns with which the occurrence of an instance of $A$ can be explained. Table~\ref{tab:lpm_instances} highlights the instances of the three LPMs in $\mathit{LPMS}$ as found by the alignment-based support scoring approach for LPMs, indicating the events that are part of an LPM instance in \textbf{bold}, and indicating a single instance of the LPM pattern by $\overline{\mbox{overline}}$. As shown earlier, LPMs \emph{(a)} and \emph{(b)} together explain all events except for the single $E$-event that is indicated in \textcolor{red}{red}. Notice that there is no overlap between LPMs \emph{(a)} and \emph{(b)} in the events that they explain: LPMs \emph{(a)} and \emph{(b)}, therefore, together they provide a near perfect explanation of the sequence database. While red $E$-event is part of a pattern instance of LPM \emph{(c)}, it cannot be explained by the LPM set, as the $D$ and $B$ events of the same pattern instance clash with an instance of LPM \emph{(a)}. One could choose to use these $D$ and $B$ events for LPM \emph{(c)} instead of LPM{(a)}, however, that would lead to two events (indicated in \textcolor{blue}{blue}) remaining unexplained for instead of only one. It is clear that there is redundancy in LPM set $\mathit{LPMS}$, as LPM \emph{(c)} does not contribute to the set of events of the sequence database that are explained by the LPMs. However, when the instances of LPMs are calculated in isolation, the degree of redundancy in a given LPM set is generally not immediately clear. 

The main problem with scoring LPMs in isolation is that an event of the sequence database can be part of a pattern instance of one LPM of the LPM set while at the same time being part of a pattern instance of another LPM. We will now propose an approach to score an LPM set such that each event \emph{can only be part of a pattern instance of one LPM}, i.e., the LPMs in the LPM set \emph{compete for the events of the sequence database} and the contribution of an LPM in the LPM set is only based on the events that it additional describes. This novel evaluation approach for sets of LPMs is based on the construction of a global model from the LPMs in the set.

\begin{table}[t]
	\centering
	\caption{An overview of the instances of the three Local Process Models in LPM set $\mathit{LPMS}$ in sequence database $\mathit{SD}$.}
	\label{tab:lpm_instances}
	\resizebox{\linewidth}{!}{
		\begin{tabular}{l|c|c|c}
			\toprule
			Sequence & LPM \emph{(a)}& LPM \emph{(b)} & LPM \emph{(c)}\\
			\midrule
			1&$\langle E,B,A,B,A,F,\overline{\textbf{A},\textbf{C},\textbf{B},\textbf{D}}\rangle$& $\langle \overline{\textbf{E},\textbf{B},\textbf{A},\textbf{B},\textbf{A},\textbf{F}},A,C,B,D\rangle$ & $\langle E,B,A,B,A,F,A,C,B,D\rangle$\\
			2&$\langle E,B,A,F,E,B,A,B,A,F\rangle$& $\langle \overline{\textbf{E},\textbf{B},\textbf{A},\textbf{F}},\overline{\textbf{E},\textbf{B},\textbf{A},\textbf{B},\textbf{A},\textbf{F}}\rangle$ & $\langle E,B,A,F,E,B,A,B,A,F\rangle$\\
			3&$\langle \overline{\textbf{A},\textbf{B},\textbf{C},\textbf{D}},\overline{\textbf{A},\textbf{C},\textbf{D},\textbf{B}},E,F\rangle$& $\langle A,B,C,D,A,C,D,B,\overline{\textbf{E},\textbf{F}}\rangle$ & $\langle A,B,C,\overline{\textbf{D},\textbf{A}},C,\overline{\textbf{D},\textbf{B},\textbf{E}},F\rangle$\\
			4&$\langle \overline{\textbf{\textcolor{blue}{A}},\textbf{\textcolor{blue}{C}},\textbf{D},\textbf{B}},\textcolor{red}{E},E,B,A,F\rangle$&$\langle \textcolor{blue}{A},\textcolor{blue}{C},D,B,\textcolor{red}{E},\overline{\textbf{E},\textbf{B},\textbf{A},\textbf{F}}\rangle$&$\langle \textcolor{blue}{A},\textcolor{blue}{C},\overline{\textbf{D},\textbf{B},\textbf{\textcolor{red}{E}}},E,B,A,F\rangle$\\
			\bottomrule
	\end{tabular}}
	\vspace{-0.3cm}
\end{table}

\subsection{Merging Local Process Models into a Global Model}
To summarize a sequence database in the form of LPMs, it is sufficient to have each event described by only one of the LPMs. To obtain an allocation of events to LPMs that provides an optimal number of explained events we transform the set of LPMs into a single process model by merging the places of the initial markings of each LPM in $\mathit{LPMS}$ into a single place $\mathit{mi}$, and set as new initial marking $\mathit{MI}=\{mi\}$ of the merged model. Furthermore, we merge the places of the final markings of the LPMs in $\mathit{LPMS}$ into a new place $\mathit{mf}$, which we set as new final marking $\mathit{MF}=\{\mathit{mf}\}$ the merged model. We will show how this global model can be used to detect instances of the LPMs in the sequence database. Formally, given an LPM set $\mathit{LPMS}=\langle\mathit{LPM}_1,\mathit{LPM}_2,\dots,\mathit{LPM}_n\rangle$ with each LPM $\mathit{LPM}_i$ being represented by an accepting Petri net $(N_i, {M_0}_i, {m_f}_i)$, with $N_i=(P_i,T_i,F_i,\ell_i)$, we first transform each Petri net $N_i$ into $N_i'=(P_i',T_i,F_i',\ell_i)$ where:
\begin{itemize}
	\item $P_i' = (P_i\cup\mathit{mi}\cup\mathit{mf})\setminus\{p|p\in {M_0}_i\cup {M_f}_i\}$
	\item $F_i' = \{(n,n')|(n,n'){\in}F_i\land n\not\in {M_0}_i{\land}n'\not\in{M_f}_i\}\cup\{(\mathit{mi},t)|(p,t){\in}F_i\land p{\in}{M_0}_i\}\cup\{(t,\mathit{mf})|(t,p){\in}F_i\land p{\in}{M_f}_i\}$
\end{itemize}

A single sequence may contain occurrences of multiple LPMs in the set. Therefore, we add a silent transition $t_\mathit{bl}$ connecting the final place to the initial place. This allows the model to accept any concatenation of occurrences of LPMs. Furthermore, it can be the case that a sequence contains no instance of any of the LPMs. Therefore, we redefine the final marking of the merged model to its initial marking to allow it to accept the empty sequence $\langle\rangle$. Formally, given an LPM set $\mathit{LPMS}{=}\langle\mathit{LPM}_1,\mathit{LPM}_2,\dots,\mathit{LPM}_n\rangle$ with each LPM $\mathit{LPM}_i$ being represented by an accepting Petri net $(N'_i, {M_0}_i, {M_f}_i)$, with $N'_i{=}(P'_i,T_i,F'_i,\ell_i)$, the merged global Petri net representing $\mathit{LPMS}$ is a Petri net $(P,T,F,\ell)$ such that:
\begin{itemize}
	\item $P=\cup_{i=1}^{|\mathit{LPMS}|}P'_i$
	\item $T=(\cup_{i=1}^{|\mathit{LPMS}|}T_i)\cup t_{\mathit{bl}}$
	\item $F=(\cup_{i=1}^{|\mathit{LPMS}|}F'_i)\cup\{(\mathit{mf},t_{\mathit{bl}}), (t_{\mathit{bl}}, \mathit{mi})\}$
	\item $\ell=\begin{cases}
	\ell_i(t),& \text{if } t\in T_i\\
	\tau,              & \text{otherwise}
	\end{cases}$
\end{itemize}

\begin{figure}
	\centering
	\includegraphics[width=0.73\linewidth]{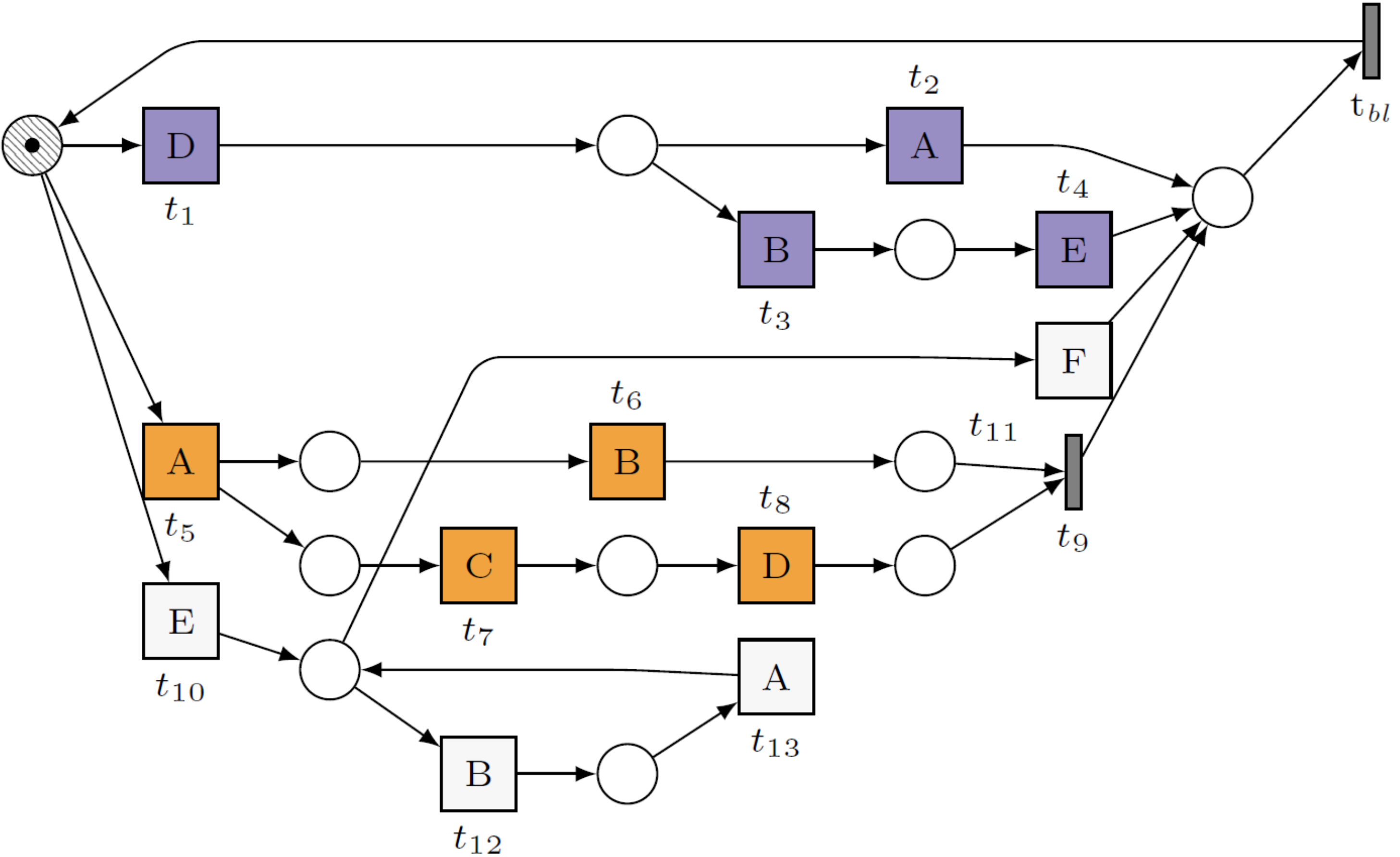}\\
	\caption{The global evaluation model for the evaluation of a Local Process Model set consisting of the Local Process Models of Figure~\ref{fig:local_process_models}.}
	\label{fig:global_merged_model}
	\vspace{-0.2cm}
\end{figure}

Figure~\ref{fig:global_merged_model} shows the global process model obtained from $\mathit{LPMS}$. \textcolor{orange}{Orange} transitions come from the LPM in Figure~\ref{fig:local_process_models}\emph{(a)}, white ones from \emph{(b)}, and \textcolor{Violet}{violet} ones from \emph{(c)}. Our approach to obtain the pattern instances of the LPMs in the LPM set based using the construction of a global model is based on a technique that is called alignments~\citep{Aalst2012}.

\subsection{Finding Instances of LPMs from an LPM Set in a Sequence Database}
The alignment algorithm~\citep{Aalst2012} calculates the best possible explanation of a sequence from a sequence database in terms of process steps in a process model. An alignment between a sequence and a process model is a pairwise matching between events and activities allowed by the model. Sometimes, events cannot be matched to any of the transitions. For instance, an event occurs when not allowed according to the model (i.e., it is not enabled). In this case, the event cannot be matched to a transition firing, resulting in so-called \emph{moves in log}. Other times, an activity should have been executed according to the model but is not observed in the sequence database. This results in a transition that cannot be matched to an event in the sequence database, thus resulting in a so-called \emph{move in model}. When an event in the sequence database can be correctly matched to a transition firing in the process model, this is called a \emph{synchronous move}. An optimal alignment \citep{Aalst2012} can be computed by searching for the mapping between a process model and a sequence that minimizes the number of moves in model and log and optimizes the number of synchronous moves, using the A$^*$ search algorithm.\looseness=-1

We can calculate the degree to which the LPM set explains the $\mathit{SD}$ using alignments on the global model constructed from the LPM set $\mathit{LPMS}$ and sequence database $\mathit{SD}$. The events that are \emph{explained} by $\mathit{LPMS}$ are the ones that are mapped to a \emph{synchronous move} in the alignments, while the unexplained events correspond to \emph{moves in log}. However, we want to count \emph{exact} and \emph{complete} observations of the LPMs, while the \emph{moves on model} option in the alignment search space allows for observations of LPM instances where activities are missing. To prevent that we count events that together only approximately (but not completely) form an execution of an LPM as being an instances of that LPM, we calculate alignments in such a way that we enforce each activity in the pattern instance to be represented by an event, i.e., we do not allow for \emph{moves on model}. Therefore, we calculate alignments where we only allow \emph{synchronous moves} and \emph{moves in log}, however, we do additionally allow for \emph{moves in model} in the specific case of silent transitions, as those transitions can only be fired through a \emph{move in model} as they have no corresponding events is the sequence database. This is however no a limitation, as silent transitions are only in the model for routing purposes and do not describe any activity. 

Table~\ref{tab:alignment} shows the alignment of the first sequence of the sequence database, $\langle A,C,D,B,E,E,B,A,F\rangle$, on the global Petri net that we constructed from the LPM set shown in Figure~\ref{fig:global_merged_model}. The alignment starts with a synchronous move on activity $A$, which the model can mimic by firing $t_5$ (enabled in the initial marking). After that, the alignment likewise can perform synchronous moves on activities $C$, $D$, and $B$. Finally, to complete one instance of LPM \emph{(a)}, silent transition $t_9$ is fired to join the two parallel branches. The sequence database cannot mimic $t_9$, leading to a model move, which is allowed since the transition is silent. Then, a model move is performed on silent transition $t_{14}$, leading to the final marking, where the process model could stop moving. However, the sequence database contains another instance of an LPM. The sequence database continues with two $E$-events, however, the process model has no way to mimic this with two consecutive firings of $E$, therefore it performs a synchronous move on one of the two $E$-events and a model move on the other one. The choice is arbitrary which $E$-event to consider for the synchronous move and which one for the model move, either choice leads to an optimal alignment. After that, synchronous moves on $B$, $A$, and $F$ bring the model to the final marking, where it can end.

The alignment of $\mathit{SD}$ on the constructed global model allows us to lift the segmentation of $\mathit{SD}$ from a single LPM to an LPM set $\mathit{LPMS}$, segmenting $\sigma\in\mathit{SD}$ into $\lambda_{1}\cdot\gamma_1^{j_{k_1}}\cdot\lambda_{2}\cdot\gamma_2^{j_2}\cdot\dotso\cdot\lambda_n\cdot\gamma_n^{j_n}\cdot\lambda_{n+1}$ such that $\gamma_i^{j_i}\in\Lan(\mathit{LPMS}(j_i))$ and $\lambda_i\not\in\cup_{\mathit{LPM}\in\mathit{LPMS}}\Lan(\mathit{LPM})$. Segmentation function $\Gamma$ lifted to LPM sets is defined as $\Gamma_{\mathit{LPMS}}(\sigma)=\gamma_1^{j_1}\cdot\gamma_2^{j_2}\cdot\dotso\cdot\gamma_n^{j_n}$ with $\gamma_i^{j_i}\in\Lan(\mathit{LPMS}(j_i))$, recognizing each event in $\sigma$ either as part of an instance of LPM $j_i$, or leaving it unexplained (i.e., the $\lambda$-segments). We again lift $\Gamma_{\mathit{LPMS}}$ to sequence databases, $\Gamma_{\mathit{LPMS}}(\mathit{SD})=\{\Gamma_{\mathit{LPMS}}(\sigma)|\sigma{\in}\mathit{SD}\}$. Furthermore, we use $\mathit{\Gamma}^{\mathit{j}}_{\mathit{LPMS}}(\mathit{SD})$ to denote the set of $\gamma$-segments that are assigned to LPM $\mathit{LPMS}(j)$. In the alignment of Table~\ref{tab:alignment} we can clearly see how alignments segment the sequence into LPM instances: the model-row of the alignment starts with transitions $t_5$,$t_7$,$t_8$, and $t_6$, which originate from the LPM of Figure~\ref{fig:local_process_models}\emph{(a)}, then it has an unmapped $E$ in the sequence-row, and then continues with transitions $t_{10}$, $t_{12}$, $t_{13}$, and $t_{11}$, which originate from the LPM of Figure~\ref{fig:local_process_models}\emph{(b)}. This indicates that alignments have segmented the sequence into first an instance ($\gamma$-segment) of the LPM of Figure~\ref{fig:local_process_models}\emph{(a)}, then a $\lambda$-segment containing $E$, and then an instance ($\gamma$-segment) of the LPM of Figure~\ref{fig:local_process_models}\emph{(b)}.

\subsection{Measuring Coverage and Redundancy of LPM Sets}
We now continue by proposing quality measures based on the segmentation of a sequence database in terms of instances of LPMs in a LPM set. First, we define \emph{coverage}, representing the ratio of events that can be explained by one of the LPMs in an LPM set $\mathit{LPMS}$:

$\mathit{coverage}(\mathit{SD},\mathit{LPMS})=\frac{\sum_{\sigma'\in\Gamma_{\mathit{LPMS}}(\mathit{SD})}|\sigma'|}{\sum_{\sigma{\in}\mathit{SD}}|\sigma|}$.

\noindent For example, the coverage of our example sequence database $\mathit{SD}$ on the example LPM set $\mathit{LPMS}$ of Figure~\ref{fig:local_process_models} is $\frac{38}{39}$, due to the one unexplained $E$-event. 

While coverage measures the share of events of $\mathit{SD}$ that is explained by $\mathit{LPMS}$, it does not measure the \emph{redundancy} of $\mathit{LPMS}$. \emph{Escaping edges precision}~\citep{Munoz2010} is a widely accepted precision measure in the field of process mining, which quantifies how much of the behavior that is allowed by the process model fits the behavior that was seen in the sequence database. Escaping edges precision is $0$ when the process model allows for all behavior over the process activities, while it is $1$ when it allows for exactly the behavior over the process activities that was seen in the sequence database. Conceptually, escaping edges precision is defined based on the states in a process model $M$ that are reached when replaying the behavior seen in $\mathit{SD}$. From these states, it is determined which possible next activity can be performed according to $M$ (called the \emph{edges}), and which of those possible next activities were from this state \emph{actually never seen} in $\mathit{SD}$ (called the \emph{escaping edges}). Escaping edges precision is inversely proportional to the number of escaping edges with respect to the total number of edges. Escaping edges precision requires the sequence database to be completely fitting on the process model. When we want to calculate the precision of the global model constructed from $\mathit{LPMS}$ with respect to $\mathit{SD}$, this is not guaranteed that this fitness requirement holds, however, we do have this guarantee when calculating the precision of the model on $\Gamma_{\mathit{LPMS}}(\mathit{SD})$ instead of on $\mathit{SD}$ itself. 

When we calculate escaping edges precision of the global model constructed from LPM set $\mathit{LPMS}$ with respect to $\Gamma_{\mathit{LPMS}}(\mathit{SD})$, the obtained value depends on several aspects of $\mathit{LPMS}$. We now summarize these aspects in three observations:

\begin{description}
	\item[\textbf{Observation 1}]{the precision measure punishes the presence of unnecessary LPMs in $\mathit{LPMS}$, i.e., LPMs that model behavior that was not seen in $\mathit{SD}$. Unnecessary LPMs lead to lower precision, as they create additional escaping edges from the start state (i.e., the initial marking).}
	\item[\textbf{Observation 2}]{the precision measure penalizes overlap in behavior between multiple LPMs in $\mathit{LPMS}$. To see that this is indeed the case, it is important to note that the implementation of $\Gamma$ deterministically maps behavior in $\mathit{SD}$ states in the model. Therefore, when LPMs $\mathit{LPM}_1$ and $\mathit{LPM}_2$ both allow for a run $\langle A,B,C\rangle$ (but both might additionally allow for other runs), all occurrences of $\langle A,B,C\rangle$ in $\mathit{SD}$ are either mapped to $\mathit{LPM}_1$ or all occurrences are mapped to $\mathit{LPM}_2$, but not a mix of both. As a result, the overlap in behavior leads either to an escaping edge in $\mathit{LPM}_1$ or in $\mathit{LPM}_2$.}
	\item[\textbf{Observation 3}]{the precision measure penalizes LPMs that, considered individually, capture too much unobserved behavior.}
\end{description}

These three observations correspond to three types of redundancy that might be present in an LPM set. The first observation corresponds to redundancy that is caused by patterns that represent behavior that does not occur in the sequence database. The second observation corresponds to redundancy that is caused by multiple patterns that (partly) model the same behavior. The third observation corresponds to redundancy within a pattern, i.e., a pattern that models behavior that does not occur in the sequence database. Given these three observations we can state that precision calculated on the global model constructed from a set of LPMs can be used to measure the degree of redundancy of that set. Note however that LPM sets without redundancy, i.e., all patterns occur in the sequence database, no behavior is modeled more than once, and the patterns do not allow for more behavior than what is seen in their instances, will yield a precision of $1$. Therefore, precision is inversely related to redundancy, and hence it can be considered to be a measure of non-redundancy.

Since we are interested in LPM set that have both high \emph{coverage} and low \emph{redundancy}, we propose to additionally measure the quality of LPM sets using \emph{F-score}: i.e., the harmonic mean of coverage and non-redundancy.

\begin{table}
	\centering
	\caption{An optimal alignment of sequence $\langle A,C,D,B,E,E,B,A,F\rangle$ on the global model of Figure~\ref{fig:global_merged_model}, as obtained with the alignment approach of \citet{Aalst2012}.}
	\label{tab:alignment}
	\scalebox{0.85}{
		\begin{tabular}{|l|c|c|c|c|c|c|c|c|c|c|c|}
			\toprule
			Log  &A    &C    &D    &B    &$\gg$ &$\gg$   &E    &E       &B       &A       &F       \\
			\midrule
			Model&A    &C    &D    &B    &$\tau$&$\tau$  &$\gg$&E       &B       &A       &F       \\
			&$t_5$&$t_7$&$t_8$&$t_6$&$t_9$ &$t_{\mathit{bl}}$&&$t_{10}$&$t_{12}$&$t_{13}$&$t_{11}$\\
			\bottomrule
	\end{tabular}}
	\vspace{-0.1cm}
\end{table}

\section{Local Process Model Collection Mining Approaches}
\label{sec:mining_approaches}
A straightforward approach to post-process the output of the basic LPM mining technique into a smaller set of patterns is to select only those LPMs that actually occur in the sequence database $\mathit{SD}$ according to the evaluation framework in Section~\ref{sec:quality_criteria} . 
To do so, we align the global model constructed from the LPMs in an LPM set $\mathit{LPMS}$ to $\mathit{SD}$ and filter out any LPM that has no instances in the sequence database. Algorithm~\ref{alg:alignment_based_selection} shows the procedure for this LPM set post-processing. We will refer to this approach as the \emph{alignment-based selection} of LPMs. Applying this filter to the LPM set shown in Figure~\ref{fig:local_process_models} and the example sequence database would result in LPM~\emph{(c)} getting filtered out of the LPM set, as it does not have instances in $\mathit{SD}$.\looseness=-1

There may exist more than one optimal alignment. For example, sequence $\sigma_1{=}\langle E,B,A,F,B,C,D\rangle$, can be aligned such that $\langle \overline{\textbf{E},\textbf{B},\textbf{A},\textbf{F}},\allowbreak B,C,D\rangle$ is one instance of LPM~\emph{(b)}, while alternatively it can be aligned such that $\langle E,B,\overline{\textbf{A},F,\textbf{B},\textbf{C},\textbf{D}}\rangle$ is one instance of LPM~\emph{(a)}, as both alignments provide an explanation for 4 out of the 7 events. When multiple optimal alignments exist, the alignment algorithm is deterministic in which optimal alignments it returns, i.e., identical sequences $\sigma_1=\sigma_2$ are always aligned to identical sequences of occurrences of LPMs $\lambda_{1}\cdot\gamma_1^{j_{k_1}}\cdot\lambda_{2}\cdot\gamma_2^{j_2}\cdot\dotso\cdot\lambda_n\cdot\gamma_n^{j_n}\cdot\lambda_{n+1}$ such that all segments $\gamma_i^{j_i}\in\Lan(\mathit{LPMS}(j_i))$ are assigned to LPM $j_i$, even when there exists an alternative LPM $\mathit{LPM}'\in\mathit{LPMS}$ with $\mathit{LPM}'\ne\mathit{LPMS}(j_i)$ such that $\gamma_i^{j_i}\in\mathit{LPM}'$. Therefore, \emph{alignment-based selection} will select only one of such LPMs to represent $\gamma_i^{j_i}$, thereby reducing the number of instances of $\mathit{LPM}$, and potentially removing it if it has no instances left, thereby reducing the redundancy in the LPM set.\looseness=-1
\newlength{\textfloatsepsave} \setlength{\textfloatsepsave}{\textfloatsep}
\setlength{\textfloatsep}{10pt}
\begin{algorithm}[t]
	\caption{Alignment-based LPM selection.}
	\label{alg:alignment_based_selection}
	\begin{algorithmic}[1]
		\renewcommand{\algorithmicrequire}{\textbf{Input:}}
		\renewcommand{\algorithmicensure}{\textbf{Output:}}
		\REQUIRE sequence database $\mathit{SD}$, LPM set $\mathit{LPMS}$
		\ENSURE  filtered LPM set $\mathit{LPMS}'$
		\\ \textit{Initialisation} :
		\STATE $i=1$
		\STATE $\mathit{LPMS}'=\langle \rangle$
		\\ \textit{Main Procedure:}
		\WHILE {$i\le|\mathit{LPMS}|$}
		\IF{$\{e\in\sigma|\sigma\in\Gamma^{i}_{\mathit{LPMS}}(\mathit{SD})\}\ne\emptyset$}
		\STATE $\mathit{LPMS}' = \mathit{LPMS}'\cdot\langle\mathit{LPMS}(i)\rangle$
		\ENDIF
		\STATE $i = i+1$
		\ENDWHILE
		\RETURN $\mathit{LPMS}'$
	\end{algorithmic}
\end{algorithm}

However, given two different sequences $\sigma_1$ and $\sigma_2$ ($\sigma_1\ne\sigma_2$) with $\mathit{hd}^k(\sigma_1)=\mathit{hd}^k(\sigma_2)$ for some prefix length $k$, there is no guarantee that the events of $\mathit{hd}^k(\sigma_1)$ and $\mathit{hd}^k(\sigma_2)$ are assigned to instances of the same LPMs. To see that this can cause redundancy to remain in the LPM set, consider $\sigma_1=\langle E,B,A,F,B,C,D\rangle$ and $\sigma_2=\langle E,B,A,F,B,C,D,A\rangle$, where $\mathit{hd}^7(\sigma_1)=\mathit{hd}^7(\sigma_2)$. As shown, for $\sigma_1$ there are two possible optimal alignments: as an instance of LPM~\emph{(a)} or as an instance of LPM~\emph{(b)}. However, $\sigma_2$, has only one optimal alignment, which is the following, where the $\langle \overline{E,B,A,F},B,C,\overline{D,A}\rangle$, with $\langle E,B,A,F\rangle$ an instance of LPM~\emph{(b)} and $\langle D,A\rangle$ an instance of LPM~\emph{(c)}. However, if we would be mining from a sequence database $\mathit{SD}=[\sigma_1,\sigma_2]$, the possible optimal alignment of $\sigma_1$ to LPM~\emph{(a)} would result in the alignment-based selection to create a redundant set of three LPMs consisting of LPMs \emph{(a)}, \emph{(b)}, and \emph{(c)}, while the alignment of $\sigma_1$ to LPM~\emph{(b)} results in an LPM set consisting of only \emph{(b)} and \emph{(c)}.

To alleviate this cause of redundancy, a greedy approach to post-process an LPM set is proposed in Algorithm~\ref{alg:greedy_LPM_selector}. The intuition behind this algorithm is as follows: first, we select the LPM from the set $\mathit{LPMS}$ that explains the highest number of events in the sequence database $\mathit{SD}$. Then, we filter out all the events from sequence database $\mathit{SD}$ that are already explained, resulting in a new sequence database $\mathit{SD}'$. Iteratively, we search for the LPMs that explain the highest number of events that were still unexplained (i.e., are in $\mathit{SD}'$), and update $\mathit{SD}'$. We call this approach introduced by Algorithm~\ref{alg:greedy_LPM_selector} the \emph{greedy selection} approach. The computational complexity of this algorithm is $\mathcal{O}(n^2)$ with $n$ the number of starting patterns, as after each step in which one pattern is selected, all the other patterns that have not yet been selected need to be considered for the next selection step. 

\begin{algorithm}[t]
	\caption{Greedy alignment-based LPM selector.}
	\label{alg:greedy_LPM_selector}
	\begin{algorithmic}[1]
		\renewcommand{\algorithmicrequire}{\textbf{Input:}}
		\renewcommand{\algorithmicensure}{\textbf{Output:}}
		\REQUIRE sequence database $\mathit{SD}$, LPM set $\mathit{LPMS}$
		\ENSURE  filtered LPM set $\mathit{LPMS}'$
		\\ \textit{Initialisation} :
		\STATE $\mathit{SD}'=\mathit{SD}$, $\mathit{LPMS}'=\langle \rangle$, $\mathit{candidate\_LPMS}=\mathit{LPMS}$, $\mathit{continue\_search}=\mathit{TRUE}$
		\\ \textit{Main Procedure:}
		\WHILE {$\mathit{continue\_search}\land |\mathit{candidate\_LPMS}|>0$}
		\STATE $i=1$, $\mathit{continue\_search} = \mathit{FALSE}$, $\mathit{max\_explained}=0$, $\mathit{best\_LPM}=\mathit{null}$
		\WHILE {$i\le|\mathit{candidate\_LPMS}|$}
		\IF{$|\{e\in\sigma|\sigma\in\Gamma_{\mathit{candidate\_LPMS}(i)}(\mathit{SD}')\}|>\mathit{max\_explained}$}
		\STATE $\mathit{max\_explained} = |\{e\in\sigma|\sigma\in\Gamma_{\mathit{candidate\_LPMS}(i)}(\mathit{SD}')\}|$
		\STATE $\mathit{best\_LPM} = \mathit{candidate\_LPMS}(i)$
		\ENDIF
		\STATE $i = i+1$
		\ENDWHILE
		\IF{$\mathit{best\_LPM}\ne\mathit{null}$}
		\STATE $\mathit{continue\_search} = \mathit{TRUE}$
		\STATE $\mathit{LPMS}' = \mathit{LPMS}'\cdot\langle\mathit{best\_LPM}\rangle$
		\STATE $\mathit{SD}'= \mathit{SD}'\downharpoonleft_{\{ e\in\sigma|\sigma\in{\Gamma_{\mathit{best\_LPM}}(\mathit{SD}')}\}}$
		\STATE $\mathit{candidate\_LPMS} = \mathit{candidate\_LPMS}{\downharpoonleft}_{\{\mathit{best\_LPM}\}}$
		
		\ENDIF
		\ENDWHILE
		\RETURN $\mathit{LPMS}'$
	\end{algorithmic}
\end{algorithm}

Algorithm~\ref{alg:greedy_LPM_selector} removes LPMs from the LPM set without taking into account how much behavior the LPMs itself allow for. The selection of only a small number of LPMs that all allow for many sequences over their activities may still result in a high degree of redundancy. In Algorithm~\ref{alg:greedy_f_score_LPM_selector} we propose a \emph{direct approach} to greedily select the best combination of LPMs from the input LPM set $\mathit{LPMS}$ that leads to the highest F-score according to the evaluation framework. We call this approach the \emph{greedy selection (F-score)} method. 

Like Algorithm~\ref{alg:greedy_LPM_selector}, the computational complexity of Algorithm~\ref{alg:greedy_f_score_LPM_selector} is also $\mathcal{O}(n^2)$ with $n$ the number of starting patterns, as in both algorithms, all $n$ patterns need to be considered for selection in each iteration and there are at most $n$ possible iterations as each pattern can only be selected once. However, while both algorithms have the same worst case complexity, we expect Algorithm~\ref{alg:greedy_f_score_LPM_selector} to be slower on the average case, for two reasons. First, the process model on which the alignment needs to be calculated to evaluate the benefit of adding an LPM to the set of selected patterns grows in every iteration in Algorithm~\ref{alg:greedy_f_score_LPM_selector}, where the process model is a global model that is constructed from a number of patterns that is growing with each iteration, while the size of the process model is stable in Algorithm~\ref{alg:greedy_LPM_selector}, where it depends on the pattern under evaluation only. Secondly, the size of the sequence database is stable in Algorithm~\ref{alg:greedy_f_score_LPM_selector}, while it shrinks with every step in Algorithm~\ref{alg:greedy_LPM_selector}. 
\begin{algorithm}[t]
	\caption{Greedy F-score based LPM selector.}
	\label{alg:greedy_f_score_LPM_selector}
	\begin{algorithmic}[1]
		\renewcommand{\algorithmicrequire}{\textbf{Input:}}
		\renewcommand{\algorithmicensure}{\textbf{Output:}}
		\REQUIRE sequence database $\mathit{SD}$, LPM set $\mathit{LPMS}$
		\ENSURE  filtered LPM set $\mathit{LPMS}'$
		\\ \textit{Initialisation} :
		\STATE $\mathit{LPMS}'=\langle \rangle$, $\mathit{candidate\_LPMS}=\mathit{LPMS}$, $\mathit{best\_fscore}=0$, $\mathit{continue\_search}=\mathit{TRUE}$
		\\ \textit{Main Procedure:}
		\WHILE {$\mathit{continue\_search}$}
		\STATE $\mathit{continue\_search}=\mathit{FALSE}$, $i=1$, $\mathit{best\_LPM}=\mathit{null}$
		\WHILE {$i\le|\mathit{candidate\_LPMS}|$}
		\IF{$\mathit{Fscore}(\mathit{SD},\mathit{LPMS}'\cdot\langle\mathit{candidate\_LPMS}(i)\rangle)>\mathit{best\_fscore}$}
		\STATE $\mathit{best\_LPM} = \mathit{LPMS}(i)$
		\ENDIF
		\STATE $i = i+1$
		\ENDWHILE
		\IF{$\mathit{best\_LPM}\ne\mathit{null}$}
		\STATE $\mathit{continue\_search}=\mathit{TRUE}$
		\STATE $\mathit{LPMS}' = \mathit{LPMS}'\cdot\langle\mathit{best\_LPM}\rangle$
		\STATE $\mathit{candidate\_LPMS}$ minus $\mathit{best\_LPM}$.
		\ENDIF
		\ENDWHILE
		\RETURN $\mathit{LPMS}'$
	\end{algorithmic}
\end{algorithm}

\subsection{Re-mining of Selected Local Process Models}
Algorithms~\ref{alg:alignment_based_selection}-\ref{alg:greedy_f_score_LPM_selector} simply select a subset of LPMs $\mathit{LPMS}'$ from an initial set of LPMs $\mathit{LPMS}$, however, the LPMs in the set themselves are left unchanged, i.e., $\forall_\mathit{LPM}\in{LPMS}':\mathit{LPM}\in\mathit{LPMS}$. However, It can be the case that two LPMs $\mathit{LPM}_1,\mathit{LPM}_2\in\mathit{LPMS}'$ are overlapping in the sequences that they allow for, i.e., $\Lan(\mathit{LPM}_1)\cap\Lan(\mathit{LPM}_2)\ne\emptyset$. If such a case, even though $\mathit{LPM}_1$ and $\mathit{LPM}_2$ are both non-redundant patterns, it does indicate that part of the behavior allowed for by $\mathit{LPM}_1$ and $\mathit{LPM}_2$ is redundant. We refer to such type of redundancy as \emph{within-LPM-redundancy}, as opposed to the \emph{between-LPM-redundancy} that Algorithms~\ref{alg:alignment_based_selection}-\ref{alg:greedy_f_score_LPM_selector} aim to address. 

To mitigate \emph{within-LPM-redundancy} from a selected set $\mathit{LPMS}'$, we propose to \emph{re-mine} a process model from the set of occurrences of each LPM, by applying any existing \emph{process discovery} algorithm to the set of pattern instances of an LPM. Algorithm~\ref{alg:remining} shows the re-mining procedure. \emph{Re-mining} is orthogonal to the selection approaches of Algorithms~\ref{alg:alignment_based_selection}-\ref{alg:greedy_f_score_LPM_selector}, and can be used in combination with any LPM selection procedure. Although re-mining can be done with any process discovery algorithm, we use the Split Miner algorithm \citep{Augusto2017,Augusto2019}, which has been shown to discover precise and simple process models. The Split Miner algorithm has linear time complexity in the number of events and polynomial time complexity in the number of activities of the sequence database~\citep{Augusto2019}. Since we apply re-mining to the instances of a pattern in a sequence database, the number of activities is restricted to only those activities that occur in the pattern, which is a small number in the case of LPM patterns. Therefore, re-mining can in practice be applied efficiently.
\begin{algorithm}[t]
	\caption{Re-mining of an LPM set.}
	\label{alg:remining}
	\begin{algorithmic}[1]
		\renewcommand{\algorithmicrequire}{\textbf{Input:}}
		\renewcommand{\algorithmicensure}{\textbf{Output:}}
		\REQUIRE sequence database $\mathit{SD}$, LPM set $\mathit{LPMS}$, Process discovery method $\mathit{PD}:\mathcal{B}(\Sigma^*)\rightarrow\mathcal{M}$
		\ENSURE  re-mined LPM set $\mathit{LPMS}'$
		\\ \textit{Initialisation} :
		\STATE $i = 1$, $\mathit{LPMS}'=\langle\rangle$
		\\ \textit{Main Procedure} :
		\WHILE {$i\le|\mathit{LPMS}|$}
		\STATE $\mathit{LPMS}'=\mathit{LPMS}'\cdot \langle\mathit{PD}(\Gamma^{i}_{\mathit{LPMS}}(\mathit{SD}))\rangle$
		\STATE $i = i+1$
		\ENDWHILE
		\RETURN $\mathit{LPMS}'$
	\end{algorithmic}
\end{algorithm}
\vspace{-0.2cm}
\subsection{Mining a Local Process Model Collection Based on Sequential Pattern Mining}
\begin{algorithm}[t]
	\caption{Mining an LPM set using CloGSgrow.}
	\label{alg:cloGSgrow}
	\begin{algorithmic}[1]
		\renewcommand{\algorithmicrequire}{\textbf{Input:}}
		\renewcommand{\algorithmicensure}{\textbf{Output:}}
		\REQUIRE sequence database $\mathit{SD}$, support threshold $\mathit{min\_sup}$, distance threshold $\mathit{min\_dist}$, distance measure $\mathit{dist}:\Sigma\time\Sigma\rightarrow[0,1]$, Process discovery method $\mathit{PD}:\mathcal{B}(\Sigma^*)\rightarrow\mathcal{M}$
		\ENSURE  LPM set $\mathit{LPMS}$
		\\ \textit{Initialisation} :
		\STATE $\mathit{clusters}=\langle\rangle$, $\mathit{explained\_events}=\emptyset$, $\mathit{LPMS}=\langle\rangle$ 
		\\ \textit{Main Procedure} :
		\STATE $\mathit{seq\_patterns}=\mathit{CloGSgrow(\mathit{SD},\mathit{min\_sup})}$
		\WHILE {$i\le|\mathit{seq\_patterns}|$}
		\IF{$(\mathit{seq\_patterns(i)}_\mathit{PI}\setminus\mathit{explained\_events})\ne\emptyset$}
		\STATE $j=1$, $\mathit{closest\_clus\_dist}=\infty$, $\mathit{closest\_clus\_ind}=\infty$
		\STATE $\mathit{explained\_events} = \mathit{explained\_events}\cup\mathit{seq\_patterns(i)}_\mathit{PI}$
		\WHILE {$j\le|\mathit{clusters}|$}
		\STATE $\mathit{min\_clus\_dist}=\infty$, $k=1$
		\WHILE {$k\le|\mathit{clusters}(j)|$}
		\STATE$\mathit{min\_clus\_dist} = \mathit{min}(\mathit{min\_clus\_dist},\mathit{dist}(\mathit{seq\_patterns}(i),[\mathit{clusters}(j)](k)))$
		\ENDWHILE
		\IF{$\mathit{min\_clus\_dist}<\mathit{closest\_clus\_dist}$}
		\STATE $\mathit{closest\_clus\_dist}=\mathit{min\_clus\_dist}$
		\STATE $\mathit{closest\_clus\_ind} = j$
		\ENDIF
		\STATE $j = j+1$
		\ENDWHILE
		\IF{$\mathit{closest\_clus\_dist}<\mathit{min\_dist}$}
		\STATE $\mathit{clusters}(\mathit{closest\_clus\_ind})=\mathit{clusters}(\mathit{closest\_clus\_ind})\cup\{\mathit{seq\_patterns(i)}\}$
		\ELSE
		\STATE 	$\mathit{clusters}=\mathit{clusters}\cdot\langle\{\mathit{seq\_patterns(i)}\}\rangle$
		\ENDIF
		\ENDIF
		\STATE $i = i+1$
		\ENDWHILE
		\STATE$i=0$
		\WHILE{$i\le|\mathit{clusters}|$}
		\STATE$\mathit{LPMS} = \mathit{LPMS} \cdot \langle\mathit{PD}(\{\mathit{SP}_\mathit{PT}|\mathit{SP}\in\mathit{clusters}(i)\})\rangle$
		\STATE $i = i+1$
		\ENDWHILE
		\RETURN $\mathit{LPMS}$
	\end{algorithmic}
\end{algorithm}
\noindent\citet{Ding2009} proposed the \emph{CloGSgrow} algorithm to mine closed repetitive gapped sequential patterns from a sequence database. This technique shares several properties with LPM discovery, which also counts the support of a pattern in a way that is repetitive (i.e., a pattern can occur multiple times per sequence) and gapped (i.e., a pattern instance does not have to be a consecutive subsequence of events). Furthermore, both techniques share the property that instances of the patterns have to be \emph{non-overlapping}, i.e., each event is part of at most one instance of the pattern. As an alternative to the introduced approaches to mine an LPM set by post-processing LPM mining results, we explore an approach to combine multiple sequential patterns together to form more complex, non-sequential LPMs. Function $\mathit{CloGSgrow(\mathit{SD},\mathit{min\_sup})}$ returns a list of closed repetitive gapped sequential patterns ordered in decreasing order by their support. Each sequential pattern $\mathit{SP}$ in the list is represented by a tuple $\langle \mathit{PT}, \mathit{PI}\rangle$, where $\mathit{PT}{\in}\Sigma^*$ represents the sequence of the pattern and $\mathit{PT}$ represents the set of events of $\mathit{SD}$ that are part of a pattern instance of $\mathit{SP}$. 
$\mathit{SP}_{\mathit{PT}}$ refers to the sequence of sequential pattern $\mathit{SP}$ and $\mathit{SP}_{\mathit{PI}}$ refers to its instances. Algorithm~\ref{alg:cloGSgrow} describes such an LPM mining approach that relies on mining closed repetitive gapped sequential patterns with the CloGSgrow algorithm, and then merges the most similar patterns using a process discovery algorithm. First, an empty set of patterns $\mathit{LPMS}$ is initialized. The algorithm first removes all patterns from the CloGSgrow patterns that do not at least explain one event that was not already explained by one of the patterns with more support. The algorithm then clusters together CloGSgrow patterns that are similar in the events of $\mathit{SD}$ that they describe, and then applies a process discovery technique to find a generalizing representation for each set of sequential patterns in the form of a process model by applying a process discovery technique. A distance measure $\mathit{dist}$ is used for the clustering of sequential patterns, where a pattern is clustered together with another pattern if their distance is less than or equal to $\mathit{max\_dist}$. In practice, we propose to use the Jaccard-distance for $\mathit{dist}$, measured between the sets of events of the sequence database that two sequential patterns describe.

The computational complexity of Algorithm~\ref{alg:cloGSgrow} is $\mathcal{O}(n)$ in the number of patterns that are found by the CloGSgrow algorithm in order to cluster them based on their similarity. Once the patterns have been clustered, process discovery is applied to each of them, which is polynomial in the number of activities. The computational complexity of this step therefore is dependent on the result of the clustering step, as the number of times that process discovery needs to be applied depends on the number of clusters and the computation time of each time that process discovery is applied depends on the number of activities that occurs in the CloGSgrow patterns in that cluster.
\subsection{Implementation}
We have implemented all algorithms for mining LPM sets that are introduced in this section, as well as the evaluation approach for LPM sets introduced in Section~\ref{sec:quality_criteria}. All algorithms and techniques are openly available as part of the process mining tool ProM~\cite{Dongen2005} in the package \emph{LocalProcessModelConformance}\footnote{\url{https://svn.win.tue.nl/repos/prom/Packages/LocalProcessModelConformance/}}.
\setlength{\textfloatsep}{\textfloatsepsave}
\vspace{-0.2cm}

\section{Evaluation}
\label{sec:experiments}
In this section, we evaluate and compare methods for mining LPM sets. Section~\ref{ssec:experimental_setup} introduces the experimental setup and in Section~\ref{ssec:experimental_results} we discuss the quantitative results of these experiments. Finally, in Section~\ref{ssec:case-study} we present and discuss the patterns that we mined from one of the datasets of the evaluation.
\subsection{Experimental Setup}
\label{ssec:experimental_setup}
In this evaluation, we compare algorithms 1, 2, 3, and 5 as introduced in Section~\ref{sec:mining_approaches} against four baseline techniques using a collection of real-life sequences databases. Furthermore, we explore the effect of the re-mining approach of algorithm 4. We now continue by detailing the baseline methods, the datasets used for the evaluation, and the evaluation methodology.

\subsubsection{Baseline Methods}
As first baseline approach we simply selects the top-k LPMs that are discovered by the LPM discovery algorithm as an LPM set. Comparison with this baseline gives insight into the effectiveness of the proposed LPM set mining techniques in reducing the redundancy from the originally mined set of LPMs. For sequence databases with more than 14 activities we use an approximate heuristic LPM mining technique proposed in \citep{Tax2016c} for computational reasons. To evaluate the quality of a set of LPMs we need to calculate alignments, as discussed in Section~\ref{sec:quality_criteria}. LPM mining can result in many thousands of patterns and the computation of alignments on a large global model that is constructed from so many patterns can become computationally infeasible. Since LPM patterns are ranked on a set of quality criteria such a \emph{support} and \emph{confidence}, we restrict the evaluation to the top 250 LPMs to make it computationally feasible to evaluate the pattern set. 

A heuristic approach to diversify the set of mined LPMs in terms of the alphabet of activities that the LPMs describe is described by \citet{Mannhardt2017}. This approach starts by selecting the top LPM from the ranking of LPMs obtained by the original LPM mining procedure, and then iterates over the ranking of LPMs, thereby selecting each LPM where the minimal Jaccard-distance of the alphabet of activities in the LPM with the alphabet of activities in one of the already selected LPMs exceeds a minimum \emph{diversity threshold}. We use the set of diversified LPMs obtained with this heuristic approach as a second baseline. This approach is solely based on the set of activities that are included in the LPMs, and in contrast to the approaches introduced in this paper does not consider the control-flow properties of the LPM. Therefore, we expect this approach to be insufficient to reduce the redundancy of a set of LPMs. However, it is computationally efficient: it is $\mathcal{O}(n)$ in the number of starting patterns.

As third baseline, we create an LPM set consisting of a single process model discovered with a traditional process discovery technique $\mathit{PD}:\mathcal{B}(\Sigma^*){\rightarrow}\mathcal{M}$. For a given sequence database $\mathit{SD}$, this creates an LPM set $\mathit{LPMS}{=}\{\mathit{PD}(\mathit{SD})\}$. We use two variants: one where we apply the Inductive Miner \citep{Leemans2013} algorithm $\mathit{PD}$, and one we apply the Split Miner algorithm~\citep{Augusto2017,Augusto2019}. A comparison with this baseline gives insight in when the mining of \emph{local patterns} is favorable instead of mining a single global model. Both process discovery algorithms are polynomial in the number of activities in the sequence database.

As fourth and final baseline we compare our approaches with the sequential patterns that we obtain with CloGSgrow~\citep{Ding2009} without using the merging approach of Algorithm~\ref{alg:cloGSgrow}. In order to compare sequential patterns with LPMs, we interpret each sequential pattern as if it is an LPM, i.e., we transform the sequential pattern into a strictly sequential Petri net where the transitions from left to right are labeled according to the sequential pattern.

\subsubsection{Evaluation Datasets}
\begin{table}
	\centering
	\caption{An overview of the sequence databases used in the experiments.}
	\label{tab:event_logs}
	\resizebox{1.03\linewidth}{!}{
		\begin{tabular}{|l|l|l|c|r|r|r|r|}
			\toprule
			ID&Name & Source & Category & \# sequences & \# events & \# activities & Perplexity\\
			\midrule
			1&BPI'12 & van Dongen\tablefootnote{\url{https://doi.org/10.4121/uuid:3926db30-f712-4394-aebc-75976070e91f}} & Business & 13087 & 164506 & 23 & 2.79\\
			2&SEPSIS & \citet{Mannhardt2016} & Business & 1050 & 15214 & 16 & 3.81\\
			3&Traffic Fine & de Leoni \& Mannhardt\tablefootnote{\url{https://doi.org/10.1007/s00607-015-0441-1}} & Business &150370&561470& 11 & 1.54\\
			4&MIT B & \citet{Tapia2004} & Human behavior & 17 & 1962 & 68 & 10.27\\
			5&Ordonez A & \citet{Ordonez2013} & Human behavior & 15 & 409 & 12 & 4.62\\
			6&Ordonez B & \citet{Ordonez2013} & Human behavior & 22 & 2334 & 12 & 4.18\\
			7&van Kasteren & \citet{Kasteren2008} & Human behavior & 23 & 220 & 7 & 3.46\\
			8&Cook hh102 labour & \citet{Cook2013} & Human behavior & 36 & 576 & 18 & 4.55\\
			9&Cook hh102 weekend & \citet{Cook2013} & Human behavior & 18 & 210 & 18 & 5.14\\
			10&Cook hh104 labour & \citet{Cook2013} & Human behavior & 43 & 2100 & 19 & 6.58\\
			11&Cook hh104 weekend & \citet{Cook2013} & Human behavior & 18 & 864 & 19 & 5.68\\
			\bottomrule
	\end{tabular}}
	\vspace{-0.3cm}
\end{table}
We perform experiments on a set of eleven real-life sequence databases, consisting of three sequence databases from the \emph{business process management} domain and eight sequence databases originating from smart home environments. Mining process model descriptions of daily life from smart home sequence databases is a novel application of process mining that has recently gained popularity \citep{Leotta2015,Sztyler2015,Tax2017b,Tax2018b}. Event data from human behavior has a high degree of variability, which has the effect that traditional process discovery methods that aim to discover a single global process model generate an overgeneralizing model, which motivates the mining of local models instead of a single global from such sequence databases.

Table~\ref{tab:event_logs} provides an overview of the eleven sequence databases that we include in the experiments and lists their size in terms of the number of sequences, events, and activities. Furthermore, the table lists the \emph{perplexity} of each sequence database as a measure of the degree of variability (i.e., the randomness) of the behavior in the sequences. The perplexity is calculated using a first order Markov model that is fitted on the sequence database, i.e., if the next activity of sequence element $t+1$ can be accurately predicted from the activity of sequence element $t$, then the perplexity is low. The perplexity is the exponentiation of the entropy and offers an intuitive interpretation: if the perplexity is $k$, then the uncertainty is equal to the roll of a $k$-sided dice.\looseness=-1

\subsubsection{Evaluation Methodology}
\label{sssec:eval-methodology}
We measure the coverage, the non-redundancy, and the F-score as introduced in Section \ref{sec:quality_criteria} for each of the techniques and on each of the datasets. We use a different minimum support for each dataset, but we keep it consistent within the dataset and thus mine CloGSgrow patterns and LPMs both with the same minimum support. Furthermore, we are interested in the \emph{complexity} of the resulting LPM sets. While in the sequential pattern mining field it is common to report the complexity of the obtained result in terms of the \emph{number of patterns that are found}, this statistic is not sufficient for the case of LPM mining, since it does not take into account the complexity of the \emph{individual LPMs} in the LPM set. When patterns can contain more complex constructs, like concurrency, loops, and choices, in addition to sequential ordering, patterns themselves can become complex in the sense that certain combinations constructs can require high cognitive load to understand the pattern.\looseness=-1

To measure the complexity of a set of patterns in a way that we take into account the complexity of the individual patterns themselves, we use two measures from the business process modeling field that have been developed to measure the complexity of a business process model. The first metric, the \emph{extended Cardoso measure}~\citep{Lassen2009}, extends an earlier measure~\citep{Cardoso2005} for the complexity of control-flow graphs to Petri nets, and is based on the presence of certain splits and joins in the syntactical process definition. This measure captures the effect that the more a process model branches using either choices or into parallel paths, the harder it gets to understand the behavior modeled by the process model. The second measure, the \emph{extended cyclomatic complexity}~\citep{Lassen2009}, extends the cyclomatic metric of~\citet{McCabe1976} to Petri nets and is based on the size of the state-space of the process model. Both measures have been shown to have a relation with the understandability of a process model~\cite{Cardoso2006,Gruhn2007}. We measure the complexity on the global model that we construct from the LPM set. In addition we also report the number of patterns.
\subsection{Results}
\label{ssec:experimental_results}
\begin{table}
	\centering
	\caption{The results of the LPM collection mining methods aggregated over the eleven sequence databases (mean $\pm$ standard deviation).}
	\label{tab:aggregate_results}
	\resizebox{1.03\linewidth}{!}{
		\begin{tabular}{| l c | r r r r|}
			\toprule
			Method & Remining & Coverage & Non-Redundancy & F-score & \# Patterns\\
			\midrule
			\emph{Baseline techniques}  & & & & &\\
			\midrule
			LPM mining & &0.6971 $\pm$ 0.272& 0.0625 $\pm$ 0.018 & 0.1095 $\pm$ 0.031& 67353 $\pm$ 94804\\
			Heuristic selection & & 0.4538 $\pm$ 0.205& 0.4021 $\pm$ 0.070& 0.4023 $\pm$ 0.139& \emph{8.0909} $\pm$ 3.390\\
			Inductive Miner & & \textbf{1.0000} $\pm$ 0.000 & 0.0412 $\pm$ 0.033 &0.0788 $\pm$ 0.062& \textbf{1.000} $\pm$ 0.000\\
			Inductive Miner (20\%)& & 0.9257 $\pm$ 0.108 & 0.0693 $\pm$ 0.032& 0.1272 $\pm$ 0.056& \textbf{1.000} $\pm$ 0.000\\
			Inductive Miner (50\%)& & 0.8033 $\pm$ 0.191& 0.1083 $\pm$ 0.071 & 0.1799 $\pm$ 0.108& \textbf{1.000} $\pm$ 0.000\\
			Split Miner & & 0.3320 $\pm$ 0.304& 0.1454 $\pm$ 0.163 & 0.1881 $\pm$ 0.201& \textbf{1.000} $\pm$ 0.000\\
			CloGSgrow & & 0.9672 $\pm$ 0.086 & 0.0106 $\pm$ 0.013 & 0.0187 $\pm$ 0.006 &	4524 $\pm$ 3466\\
			\midrule
			\emph{New approaches}  & & &&&\\
			\midrule
			Heuristic selection & $\checkmark$ & 0.4538 $\pm$ 0.205& 0.4184 $\pm$ 0.064& 0.4124 $\pm$ 0.144 & \emph{8.0909} $\pm$ 3.390\\
			Alignment-based selection & & 0.6971 $\pm$ 0.272& 0.1565 $\pm$ 0.043 & 0.2354 $\pm$ 0.066&43.4545 $\pm$ 22.967\\
			Alignment-based selection & $\checkmark$& 0.6965 $\pm$ 0.272& 0.1859 $\pm$ 0.040 & 0.2687 $\pm$ 0.058& 26.6364 $\pm$ 14.438\\
			Greedy selection & & 0.5620 $\pm$ 0.230 & 0.3333 $\pm$ 0.069 & 0.3904 $\pm$ 0.108& 14.4545 $\pm$ 6.251\\
			Greedy selection & $\checkmark$ & 0.5571 $\pm$ 0.226 & 0.3618 $\pm$ 0.083 & 0.4118 $\pm$ 0.121& 14.4545 $\pm$ 6.251\\
			Greedy selection (F-score) & & 0.5063 $\pm$ 0.201 & \textbf{0.5766} $\pm$ 0.096& \textbf{0.5191} $\pm$ 0.150& 8.7273 $\pm$ 5.236\\
			Greedy selection (F-score) & $\checkmark$ & 0.4929 $\pm$ 0.189& 0.5579 $\pm$ 0.105& 0.5055 $\pm$ 0.146& 8.7273 $\pm$ 5.236\\
			CloGSgrow merging & & 0.5557 $\pm$ 0.101 & 0.4768 $\pm$ 0.256 & 0.4596 $\pm$ 0.165& 8.2727 $\pm$ 5.293\\
			\bottomrule
	\end{tabular}}
	\vspace{-0.cm}
\end{table}
Table~\ref{tab:aggregate_results} shows the mean coverage, non-redundancy, and F-score for each of the LPM set mining approaches averaged over all 11 sequence databases. It shows that LPM mining without reducing redundancy indeed results in very high numbers of patterns, results in very redundant LPM sets, and results in LPM sets that have a high coverage of $0.6971$. Likewise, the set of CloGSgrow patterns without the merging post-processing steps results in even higher coverage, high redundancy, and high numbers of patterns. As expected, the number of patterns is higher for LPM mining than for CloGSgrow pattern mining, as the non-sequential nature of the patterns allow for more variations. The fact that lower coverage and lower redundancy was found for LPM patterns than for CloGSgrow patterns is likely do to the fact that we selected only the top 250 LPM patterns for computational reasons. 

All of the proposed post-processing techniques for LPM sets and for sets of CloGSgrow patterns succeed in bringing down the redundancy and the number of patterns, but they come at the cost of coverage. The different proposed techniques differ in the trade-off between coverage and redundancy that they provide. The greedy F-score based selection approach on average results in the lowest redundancy and F-score over the 11 sequence databases, explaining on average $\pm$50\% of the events in the log with a non-redundancy of 0.5766, however this approach reduces the coverage the most with respect to the original set of LPM patterns. On the other end of the spectrum, the alignment-based selection approach does not reduce coverage at all with respect to the original set of LPMs, but it is only able to reduce redundancy to some degree. 

The original CloGSgrow patterns, which are not merged into complex non-sequential patterns using the proposed merging procedure, consistently have a high coverage and on average over all the datasets explain almost 97\% of the events. Note that this coverage depends on the minimal support that is used in the mining, lower minimum support values correspond to higher coverage values, and, in the extreme, a minimum support of only $1$ guarantees a coverage of $1.0$. However, lower minimum support also results in higher number of patterns, leading to high redundancy. The CloGSgrow merging approach creates patterns out of the CloGSgrow patterns in a way that this redundancy is substantially reduced, from an average non-redundancy of $0.0106$ to an average of $0.4768$, however the coverage is significantly reduced to $0.5557$. The reduction in coverage is caused by the application of the Split Miner~\citep{Augusto2017,Augusto2019}, which doesn't guarantee the discovery of a fitting process model, thereby reducing coverage. The LPM sets that are mined with the CloGSgrow merging procedure on average cover slightly more events than the LPM sets that are discovered with the greedy F-score based approach, however, these LPM sets are on average more redundant. The high standard deviation of the redundancy of the LPM sets obtained with the CloGSgrow merging approach shows that the approach is unstable in the quality of the results, resulting in very redundant LPM sets of some sequence databases and very non-redundant LPM sets on others.

The Inductive Miner algorithm \citep{Leemans2013} provides a formal guarantee that all behavior of the $\mathit{SD}$ is contained in the model, therefore, a coverage of 1 is guaranteed. However, it leads to very imprecise process models. When the Inductive Miner with infrequency filtering (20\% or 50\%) is used, the coverage of the resulting LPM set decreases while non-redundancy slightly increases. The Split Miner~\citep{Augusto2017,Augusto2019} outperforms the Inductive Miner in terms of redundancy, however, the coverage of the resulting process models is unstable, even resulting in a coverage of 0 on some sequence databases. The coverage values of 0 are caused by process models being generated by the Split Miner where none of the sequences of the sequence database fits on the model without the need of skipping at least one activity somewhere in the model. The alignment-based selection approach post-processes the LPM mining results leading to lower redundancy, without any loss in coverage. In contrast, the greedy approach and the greedy F-score based approach are not able to post-process LPM mining results without loss in coverage, however, those two approaches are able to obtain higher reduction of redundancy.\looseness=-1

\subsubsection{Detailed Results}
Table~\ref{tab:fscore_results} shows more fine-grained results by showing the F-score obtained with each of the LPM set mining approaches on each of the 11 sequence databases individually. While the greedy F-score based selection approach on average results in the highest F-score, the approach based on merging CloGSgrow pattern outperform this approach on two of the eleven datasets (i.e., BPI'12 and MIT B). In contrast, the merged CloGSgrow patterns perform substantially less well on the Ordonez B, the van Kasteren, and the Cook hh102 weekend datasets. We conjecture that the relative performance of the greedy F-score based selection and the CloGSgrow merging approaches are related to the length of the frequent patterns in the data set. Where the mined LPMs are restricted to at most four activities for computational reasons, the CloGSgrow patterns do not have this restriction, and the patterns that meet the support threshold can be considerably larger. Therefore, long sequences of frequently repeated behavior can be captured in a single CloGSgrow pattern, where multiple LPMs are required, leading to a lower redundancy for the merged CloGSgrow patterns compared to the LPMs selected with the greedy F-score selection. On the other hand, when there are no long frequent patterns, the maximum size restriction of LPMs does not pose a problem. This conjecture is supported by Table \ref{tab:clogsgrow_pattern_length}, which shows the pattern length of the ten longest CloGSgrow patterns per sequence database, as for two of the three sequence databases on which the CloGSgrow merging procedure outperforms the greedy F-score based approach the mined CloGSgrow patterns were substantially lower than the ones that were mined for the other sequence databases. 

\begin{table}
	\centering
	\caption{The F-score of the LPM set mining methods per sequence database.}
	\label{tab:fscore_results}
	\resizebox{1.03\linewidth}{!}{
		\begin{tabular}{ |lc | r r r r r r r r r r r|}
			\toprule
			&&\multicolumn{11}{c|}{Sequence database (IDs as shown in Table~\ref{tab:event_logs})}\\
			\cmidrule{3-13}
			Method & R & 1 & 2 & 3 & 4 & 5 & 6 & 7 & 8 & 9 & 10 & 11\\
			\midrule
			\emph{Baseline techniques}  & & & &&&&&&&&&\\
			\midrule
			LPM mining & & 0.0866 & 0.1033 & 0.0615 & 0.0937 & 0.1103 & 0.0820 & 0.0605& 0.1306 & 0.1163 & 0.1600 & 0.1521 \\
			Heuristic selection & & 0.2491 & 0.4323 & 0.5056 & 0.1263 & 0.4709 & 0.4973 & 0.3238 & 0.4429 & 0.4459 & 0.5140 & 0.4854 \\
			Inductive Miner & & 0.1713 & 0.1827 & 0.1939 & 0.0014& 0.0059 & 0.0844 & 0.1267 & 0.0315 & 0.0037 & 0.1379 & 0.1326\\
			Inductive Miner (20\%)& & 0.2184 & 0.2319 & 0.1804 & 0.0217 & 0.1003 & 0.1304 & 0.2169 & 0.1618 & 0.1204 & 0.1137 & 0.1463 \\
			Inductive Miner (50\%)& & 0.3142 & 0.2155 & 0.1669 & 0.0059 & 0.2488 & 0.3497 & 0.2303 & 0.2032 & 0.1209 & 0.0979 & 0.1329 \\
			Split Miner & & 0.0000 & 0.0000 & 0.0000 & 0.0000 & 0.3230 & 0.3004 & 0.5347 & 0.0605 & 0.2855 & 0.0074 & 0.0371 \\
			CloGSgrow & & 0.0187 & 0.0215 & 0.0236 & 0.0212 & 0.0209 & 0.0109 & 0.0205 & 0.0148 & 0.0131 & 0.0107 & 0.0301 \\	
			\midrule
			\emph{Novel approaches}  & & & &&&&&&&&&\\
			\midrule
			Heuristic selection & $\checkmark$ & 0.2491 & 0.4323 & 0.5056 & 0.1281 & 0.4709 & 0.4973 & 0.4559 & 0.5217 & 0.4459 & 0.5140 & 0.4854\\
			Alignment-based selection & & 0.2070 & 0.1863 & 0.2292 & 0.1495 & 0.3227 & 0.1369 & 0.2535 & 0.3053 & 0.2829 & 0.2194 & 0.2909\\
			Alignment-based selection & $\checkmark$& 0.2528 & 0.2323 & 0.3474 & 0.1576 & 0.3416 & 0.2234 & 0.3094 & 0.3229 & 0.3147 & 0.2319 & 0.2999\\
			Greedy selection & & 0.2451 & 0.4387 & 0.5222 & 0.1552 & 0.4275 & 0.3667 & 0.4820 & 0.4725 & 0.4414 & 0.5638 & 0.4275\\
			Greedy selection & $\checkmark$ & 0.2451 & 0.4425 & 0.5222 & 0.1555 & 0.4666 & 0.3993 & 0.5624 & 0.4863 & 0.4722 & 0.4599 & 0.4284\\
			Greedy selection (F-score) & & 0.3983 & \textbf{0.5750} & \textbf{0.6927} & 0.1777 & \textbf{0.6017} & \textbf{0.6507} & \textbf{0.6674} & \textbf{0.5670} & 0.5459 & \textbf{0.5758} & 0.5327\\
			Greedy selection (F-score) & $\checkmark$ & 0.3983 & \textbf{0.5750} & \textbf{0.6927} & 0.1763 & \textbf{0.6017} & 0.5228 & \textbf{0.6674} & \textbf{0.5670} & \textbf{0.5474} & \textbf{0.5758} & 0.5327\\
			CloGSgrow merging & & \textbf{0.4562} & 0.3785 & 0.4590 & \textbf{0.3331} & 0.5261 & 0.3212 & 0.3478 & 0.4866 & 0.1918 & 0.5433 & \textbf{0.5685}\\
			\bottomrule
	\end{tabular}}
	\vspace{-0.4cm}
\end{table}

The naive heuristic selection approach one average is outperformed in F-score only by the greedy F-score based approach and the CloGSgrow merging approach. This is remarkable, given that this approach does not take into account overlap in behavior between LPMs and merely looks at which activities are modeled. However, on specific datasets, such as the MIT B dataset, it returns a set of LPMs with considerably lower F-score than other approaches. Given that this naive approach has a very favorable runtime complexity compared to the other approaches, it can be useful as a starting point before moving to more involved but more computationally intensive techniques.

The re-mining procedure with the Split Miner algorithm decreases the redundancy and increases the F-score when applied to the results of the heuristic selection, alignment-based selection, and greedy selection. This comes at the price of a minor decrease in coverage. Surprisingly, the re-mining procedure has a negative effect on coverage, redundancy and F-score when used together with the greedy F-score based approach. Furthermore, the Inductive Miner performs much better for sequence databases 1,2,3, and 7, which are the datasets with the lowest perplexity. Furthermore, the Split Miner performs very well for sequence database 7. This shows that traditional process discovery approaches perform well when the sequence database is highly structured, while mining of \emph{local patterns} instead of a \emph{global model} performs better when there is less structure in a sequence database. 

\begin{table}
	\centering
	\caption{The sequence length of the ten longest CloGSgrow patterns per sequence database.}
	\label{tab:clogsgrow_pattern_length}
	\scalebox{0.8}{
		\begin{tabular}{ |l | r r r r r r r r r r r|}
			\toprule
			&\multicolumn{11}{c|}{Sequence database (IDs as shown in Table~\ref{tab:event_logs})}\\
			\cmidrule{2-12}
			Pattern & 1 & 2 & 3 & 4 & 5 & 6 & 7 & 8 & 9 & 10 & 11\\
			\midrule
			pattern 1 & 48 & 8 & 7 & 154 & 11 & 20 & 8 & 11 & 8 & 11 & 8 \\
			pattern 2 & 47 & 8 & 6 & 153 & 11 &19 & 8 & 11 & 8 & 11 & 8 \\
			pattern 3 & 46 & 8 & 6 & 152 & 11 & 18 & 8 & 11 & 8 & 11 & 8\\
			pattern 4 & 45 & 8 & 6 & 151 & 11 & 17 & 8 & 11 & 8 & 11 & 8 \\
			pattern 5 & 44 & 8 & 6 &150 & 11 & 16 & 8 & 11 & 8 & 11 & 8 \\
			pattern 6 & 43 & 8 & 6 &  149 & 11 & 15 & 8 & 11 & 8 & 11 & 8 \\
			pattern 7 & 42 & 8 & 6 & 148 & 11 & 14 & 8 & 11 & 8 & 10 & 8 \\	
			pattern 8 & 41 & 8 & 5 & 147 & 11 & 13 & 8 & 11 & 8 & 10 & 8 \\	
			pattern 9 & 40 & 8 & 5 &  146 & 11 & 12 & 8 & 11 & 7 & 10 & 8 \\	
			pattern 10 & 39 & 8 & 5 & 145 & 11 & 11 & 7 & 11 & 7 & 10 & 8 \\	
			\bottomrule
	\end{tabular}}
	\vspace{-0.4cm}
\end{table}

\begin{table}
	\caption{The \emph{number of LPMs} (L) / \emph{number of transitions} (T) / \emph{extended Cardoso measure} (CA) / \emph{extended cyclomatic complexity} (CY) of the LPM set mining methods per sequence database.}
	\label{tab:cyclomatic_complexity_results}
	\resizebox{1.03\linewidth}{!}{
		\subfloat{
			\begin{tabular}{ |l c | r r r r | r r r r | r r r r | r r r r |}
				\toprule
				&&\multicolumn{16}{c|}{Sequence database (IDs as shown in Table~\ref{tab:event_logs})}\\
				\cmidrule{3-18}
				Method & R & \multicolumn{4}{c|}{1} & \multicolumn{4}{c|}{2} & \multicolumn{4}{c|}{3} & \multicolumn{4}{c|}{4} \\
				\midrule
				\emph{Baseline techniques}  & & L & T & CA & CY &L& T & CA & CY&L& T & CA & CY&L& T & CA & CY\\
				\midrule
				LPM mining & &112365&1021&530&313&5919&912&462&223&306872&729&290&208&6423&493&503&226\\
				Heuristic selection &&5&17&16&8&13&44&44&19&6&18&15&9&10&37&38&177\\
				Inductive Miner &&1&55&41&44&1&33&22&48&1&25&19&67&1&92&31&1491\\
				Inductive Miner (20\%)&&1&47&34&30&1&38&29&92&1&27&25&497&1&86&41&791\\
				Inductive Miner (50\%)&&1&44&41&27&1&39&31&67&1&24&24&438&1&75&\textbf{29}&312\\
				Split Miner &&1&138&147&-&1&161&186&-&1&49&60&-&1&327&324&232\\	
				CloGSgrow &&6478&194340&-&-&4039&10893&-&-&2394&4325&-&-&12474&125728&-&-\\
				\midrule
				\emph{Novel approaches}  & & & & & & & & & & & & & & & & & \\
				\midrule
				Heuristic selection & $\checkmark$ &5&17&16&8&13&44&44&19&6&18&15&9&10&\textbf{33}&34&\textbf{15}\\
				Alignment-based selection & &16&59&64&30&61&228&233&112&24&81&67&50&45&148&149&67\\
				Alignment-based selection & $\checkmark$&14&44&46&23&50&141&114&76&17&49&45&24&35&105&106&51\\
				Greedy selection & &4&\textbf{13}&\textbf{13}&8&18&66&69&\textbf{4}&13&39&40&17&21&66&67&29\\
				Greedy selection &$\checkmark$&4&\textbf{13}&\textbf{13}&8&18&66&69&\textbf{4}&13&35&36&15&21&56&62&28\\
				Greedy selection (F-score) & &6&28&28&13&2&\textbf{9}&\textbf{10}&\textbf{4}&3&\textbf{12}&\textbf{10}&\textbf{6}&20&76&77&34\\
				Greedy selection (F-score) & $\checkmark$ &6&27&22&10&2&\textbf{9}&\textbf{10}&\textbf{4}&3&\textbf{12}&\textbf{10}&\textbf{6}&20&68&68&31\\
				CloGSgrow merging &&3& 14&\textbf{13}&\textbf{6}&9&45&38&23&8&34&29&13 &7&64&61&46\\
				\bottomrule
	\end{tabular}}}\\[1ex]
	
	\resizebox{1.03\linewidth}{!}{
		\subfloat{
			\begin{tabular}{ |l c | r r r r | r r r r | r r r r | r r r r |}
				\toprule
				&&\multicolumn{16}{c|}{Sequence database (IDs as shown in Table~\ref{tab:event_logs})}\\
				\cmidrule{3-18}
				Method & R & \multicolumn{4}{c|}{5} & \multicolumn{4}{c|}{6} & \multicolumn{4}{c|}{7} & \multicolumn{4}{c|}{8}\\
				\midrule
				\emph{Baseline techniques}  & & L & T & CA & CY &L& T & CA & CY&L& T & CA & CY&L& T & CA & CY\\
				\midrule
				LPM mining & &151624&1013&509&252&136887&1145&504&276&12036&998&431&234&2865&850&383&193\\
				Heuristic selection &&12&38&39&17&4&\textbf{13}&\textbf{14}&6&4&14&12&7&12&42&38&20\\
				Inductive Miner &&1&47&50&214&1&25&17&72&1&23&21&225&1&37&27&22\\
				Inductive Miner (20\%)&&1&31&34&1823&1&21&17&14&1&20&17&96&1&29&21&18\\
				Inductive Miner (50\%)&&1&27&28&181&1&26&24&17&1&19&18&83&1&32&\textbf{18}&22\\
				Split Miner &&1&59&60&-&1&84&84&34&1&27&27&-&1&32&22&-\\
				CloGSgrow &&6065&10896&-&-&8963&70361&-&-&333&1630&-&-&628&5089&-&-\\
				\midrule
				\emph{Novel approaches}  & & & & & & & & & & & & & & & & &\\
				\midrule
				Heuristic selection & $\checkmark$ &12&38&39&17&4&\textbf{13}&\textbf{14}&6&4&14&12&7&12&35&32&17\\
				Alignment-based selection & &43&146&144&72&91&366&323&179&20&87&83&44&33&117&114&53\\
				Alignment-based selection & $\checkmark$&16&89&87&41&55&161&161&76&14&48&48&21&23&79&78&37\\
				Greedy selection & &12&38&39&19&13&44&44&19& 7&22&19&14&18&59&61&26\\
				Greedy selection &$\checkmark$&12&30&30&16&13&41&41&18&7&21&19&12&18&54&54&24\\
				Greedy selection (F-score) & &12&41&39&21&7&23&23&10&4&\textbf{12}&\textbf{10}&\textbf{6}&10&31&30&14\\
				Greedy selection (F-score) & $\checkmark$ &12&41&39&21&7&22&23&12&4&13&12&\textbf{6}&10&31&30&14\\
				CloGSgrow merging &&6&\textbf{20}&\textbf{19}&\textbf{6}&4&15&\textbf{14}&7&5&16&16&\textbf{6}&7&\textbf{25}&24&\textbf{12}\\
				\bottomrule
	\end{tabular}}}\\[1ex]
	
	\resizebox{0.81\linewidth}{!}{
		\subfloat{
			\begin{tabular}{ |l c | r r r r | r r r r | r r r r|}
				\toprule
				&&\multicolumn{12}{c|}{Sequence database (IDs as shown in Table~\ref{tab:event_logs})}\\
				\cmidrule{3-14}
				Method & R & \multicolumn{4}{c|}{9} & \multicolumn{4}{c|}{10} & \multicolumn{4}{c|}{11} \\
				\midrule
				\emph{Baseline techniques}  & & L & T & CA & CY & L & T & CA & CY& L & T & CA & CY\\
				\midrule
				LPM mining & & 2427&831&368&203&1208&899&427&215&2256&916&413&211 \\
				Heuristic selection &&10&37&32&18&6&\textbf{19}&19&\textbf{9}&7&\textbf{22}&22&\textbf{10} \\
				Inductive Miner &&1&30&\textbf{19}&331&1&26&\textbf{11}&20&1&28&15&20\\
				Inductive Miner (20\%)&&1&29&24&107&1&38&25&23&1&24&\textbf{11}&18 \\
				Inductive Miner (50\%)&&1&\textbf{26}&27&1172&1&32&18&22&1&25&\textbf{11}&20 \\
				Split Miner&&1&59&59&-&1&264&298&-&1&119&121&- \\
				CloGSgrow &&2225&12609&-&-&3698&14931&-&-&2469&13589&-&-\\	
				\midrule
				\emph{Novel approaches} &&&&&&&&&&&&&  \\
				\midrule
				Heuristic selection & $\checkmark$ &10&35&31&17&6&\textbf{19}&19&\textbf{9}&7&\textbf{22}&22&\textbf{10}\\
				Alignment-based selection & &28&93&85&44&69&231&223&11&48&193&181&90\\
				Alignment-based selection & $\checkmark$&16&69&69&32&24&85&84&71&29&115&114&53\\
				Greedy selection  &&11&34&27&18&26&85&84&41&16&53&52&26\\
				Greedy selection &$\checkmark$&11&33&29&16&26&78&75&41&16&50&48&25\\
				Greedy selection (F-score) &&11&34&31&16&13&42&42&20&8&27&27&12\\
				Greedy selection (F-score) & $\checkmark$ &11&31&28&\textbf{15}&13&42&42&20&8&27&27&12\\
				CloGSgrow merging &&22&97&97&527&12&39&37&31&9&29&30&26\\
				\bottomrule
	\end{tabular}}}
	\vspace{-0.4cm}
\end{table}

Table~\ref{tab:cyclomatic_complexity_results} shows the patterns, the number of transitions and the complexity of the patterns in terms of extended Cardoso measure and extended cyclomatic complexity. For the LPM mining results, the number of LPMs reports the total number of patterns that are found, but the number of transitions, the extended Cardoso measure, and the extended cyclomatic complexity measure are reported on the top 250 LPMs, as discussed in Section~\ref{sssec:eval-methodology}.

The results show that all post-processing approaches for LPM mining results are effective in substantially bringing down the complexity of the LPM set, thereby creating much more understandable results for the user by reducing the pattern overload. It depends per sequence database which of the algorithms results in the most simple LPM set. It seems that while there are substantial differences between the approaches in computational complexity and in trade-off that the approaches offer between coverage and redundancy, in contrast, the differences between the approaches in pattern complexity are small and cannot be considered substantial.

Traditional process discovery techniques, compared to LPM sets, create only a single ``pattern'' and this single pattern has a low numbers of transition compared to the global model that we construct out of LPM sets that we obtain with the proposed LPM post-processing approaches. However, these models that are discovered with process discovery approaches often are more complex in terms of extended Cardoso measure and extended cyclomatic complexity, indicating that they might in fact be more difficult for the analyst to understand than the set of LPM patterns. The Split Miner generates a model with \emph{improper completion} for many of the sequence databases, i.e., it allows for runs through the model that do not end with one token in the final marking. This is widely considered to be an undesirable property of a process model, and it disallows the calculation of the extended cyclomatic complexity (resulting in "-" values in the CY column of Table~\ref{tab:cyclomatic_complexity_results}).

Note that we did not calculate complexity measures other than the number of LPMs and the number of transitions for CloGSgrow patterns. Because the patterns obtained with CloGSgrow are strictly sequential, it would not yield additional insights to calculate complexity measures for process models for them, and it suffices to compare strictly on number of patterns and transitions. Compared to LPM mining, CloGSgrow generates a smaller number of patterns on almost all datasets. This is expected, as the non-sequential constructs that are allowed in LPMs allow for a wider variation of patterns. Exception however are the Cook hh104 labour and hh104 weekend datasets, on which CloGSgrow generated more patterns than LPM mining even though the same minimal support was used for mining. The reason for this is that for computational reasons we used an approximate heuristic LPM mining approach~\citep{Tax2016c} for datasets with 15 or more activities, resulting in the situation that some LPM patterns that meet the minimum support are not found.

\subsubsection{Computation Time}
In addition to the computational complexity of the approaches that we reported in Section~\ref{sec:mining_approaches}, we now discuss experimental findings of the runtime of the approaches on the eleven datasets. Table~\ref{tab:runtime_results} shows the running time (in seconds) of each approach on each of the 11 sequence databases, on an Intel i7 CPU @ 2.4GHz with 16 GB of memory. 

The traditional process discovery approaches (the Inductive Miner and the Split Miner) are very fast and can generate a process model from the sequence database in less than a second. For techniques that post-process LPM mining results (i.e., heuristic selection, alignment-based selection, greedy selection, and greedy F-score based selection) the reported computation times include the time for mining the LPMs. LPM mining is slow for the traffic fine sequence database because this sequence database has many sequences and 11 activities, which is just below the threshold (14 activities) for switching to the heuristic approximate LPM mining approach \citep{Tax2016c}. Alignment-based selection of LPMs is often slower than LPM mining itself. Heuristic selection is a fast procedure, adding at most a second to the computation time needed for mining the LPMs.

For the CloGSgrow merging approach the computation times include only the merging procedure and not the time required to mine the CloGSgrow patterns themselves. This is because we relied on our own implementation of the CloGSgrow algorithm and want to prevent implementation-specifics from influencing the results.

\begin{table}
	\centering
	\caption{The runtime (in seconds) of the LPM set mining methods per sequence database.}
	\label{tab:runtime_results}
	\resizebox{1.03\linewidth}{!}{
		\begin{tabular}{ |l c | r r r r r r r r r r r|}
			\toprule
			&&\multicolumn{11}{c|}{Sequence database (IDs as shown in Table~\ref{tab:event_logs})}\\
			\cmidrule{3-13}
			Method & Remining & 1 & 2 & 3 & 4 & 5 & 6 & 7 & 8 & 9 & 10 & 11\\
			\midrule
			\emph{Baseline techniques}  & & & &&&&&&&&&\\
			\midrule
			LPM mining & & \emph{790} & \emph{8} & 26824 & \emph{142} & 123 & 700 & 8 & \emph{18} & \emph{11} & \emph{589} & \emph{189} \\
			Heuristic selection & & 790 & 8 & 26825 & 142 & 124 & 701 & 8 & 19 & 12 & 589 & 190 \\
			Inductive Miner & & 2 & 1 & 5 & 1 & 1 & 1 & 1 & 1 & 1 & 1 & 1\\
			Inductive Miner (20\%)& & 2 & 1 & 4 & 1 & 1 & 1 & 1 & 1 & 1 & 1 & 1 \\
			Inductive Miner (50\%)& & 1 & 1 & 3 & 1 & 1 & 1 & 1 & 1 & 1 & 1 & 1 \\
			Split Miner & & 1 & 1 & 1 & 1 & 1 & 1 & 1 & 1 & 1 & 1 & 1 \\
			\midrule
			\emph{Novel approaches} & & & &&&&&&&&&\\
			\midrule
			Heuristic selection & $\checkmark$ & 791 & 9 & 26825 & 144 & 124 & 701 & 9 & 19 & 12 & 590 & 191\\
			Alignment-based selection & & 7202 & 323 & 53149 & 5414 & 1110 & 1071 & 19 & 476 & 2058 & 8994 & 6421\\
			Alignment-based selection & $\checkmark$& 7203 & 325 & 53151 & 5421 & 1114 & 1073 & 20 & 478 & 2060 & 9001 & 6433\\
			Greedy selection & & 1542 & 841 & 28641 & 5570 & 201 & 852 & 532 & 130 & 86 & 774 & 316\\
			Greedy selection & $\checkmark$ & 1544 & 843 & 28643 & 5574 & 202 & 854 & 534 & 132 & 88 & 779 & 318\\
			Greedy selection (F-score) & & 8012 & 764 & 28757 & 6311 & 352 & 1103 & 308 & 331 & 199 & 1590 & 882\\
			Greedy selection (F-score) & $\checkmark$ & 8014 & 765 & 28760 & 6315 & 356 & 1105 & 310 & 333 & 201 & 1593 & 886\\
			CloGSgrow merging & & 1412 & 1632 & 17197 & 4983 & 1358 & 603 & 325 & 275 & 206 & 689 & 340\\
			\bottomrule
	\end{tabular}}
	\vspace{-0.3cm}
\end{table}

\subsection{Case Study}
\label{ssec:case-study}
In this section we present and discuss the set of Local Process Models that we obtained with the greedy F-score based selection approach on the BPI'12 dataset, i.e., dataset 1 from Table~\ref{tab:event_logs}. This sequence database originates from a financial loan application process at a large Dutch financial institution and every sequence represents the business activities that are performed for a single loan application.

The names of the activities in this sequence database all start with \emph{A\_}, \emph{O\_}, or \emph{W\_}. Activities starting with \emph{A\_} refer to the financial loan application themselves, i.e., they correspond to status changes of the financial loan application. Examples are \emph{A\_SUBMITTED}, which indicates that the application has been submitted, \emph{A\_PREACCEPTED}, which indicates that there is a provisional decision to accept the application but additional information from the applicant is required, and \emph{A\_ACTIVATED}, which indicates that an accepted financial loan has gone into effect and payments are being processed. Activities that start with \emph{O\_} correspond to offers that are communicated to the customer by bank employees. Examples include \emph{O\_SENT}, which indicates that a offer that was prepared by a bank employee has been sent to the customer, \emph{O\_CANCELLED}, which means that the bank employee cancelled an offer, and \emph{O\_DECLINED}, which indicates that the customer has refused an offer. Finally, activities that start with \emph{W\_} correspond to work items, i.e., tasks that are assigned to bank employees for the approval process. Examples include \emph{W\_ Afhandelen leads} (in English: ``Processing leads''), which indicates that a bank employee follows up on an incomplete submission of a loan application by a potential customer, \emph{W\_Nabellen offertes} (in English: ``Calling after offers'') which indicates a call from a bank employee after transmitting an offer to a customer, and \emph{E\_Nabellen incomplete dossiers} (in English: ``Calling after incomplete files''), which indicates a call from a bank employee to request additional information from the potential customer that is needed to asses the application.

\begin{figure}
	\centering
	\subfloat[\label{sfig:bpic-lpm_1}support: 5015]{
		\includegraphics{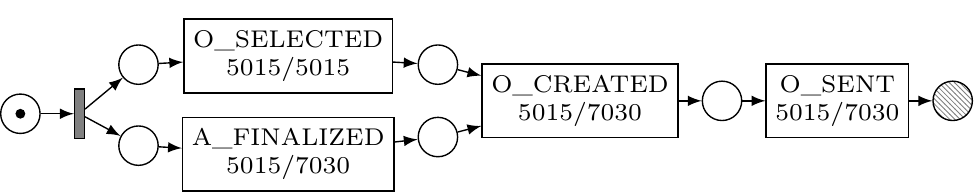}
	}
	
	\subfloat[\label{sfig:bpic-lpm_2}support: 1479]{
		\includegraphics[width=\linewidth]{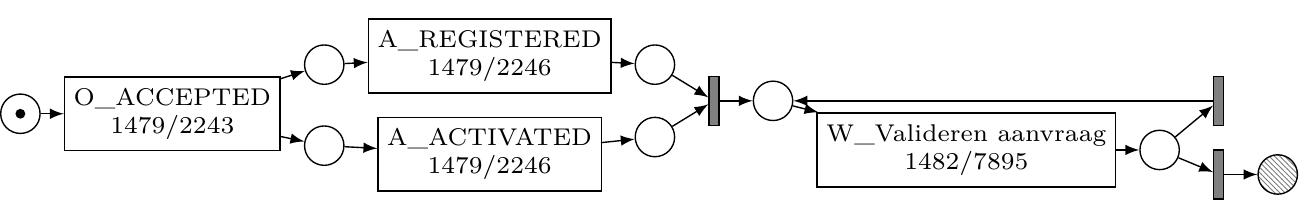}
	}
	
	\subfloat[\label{sfig:bpic-lpm_3}support: 2592]{
		\includegraphics{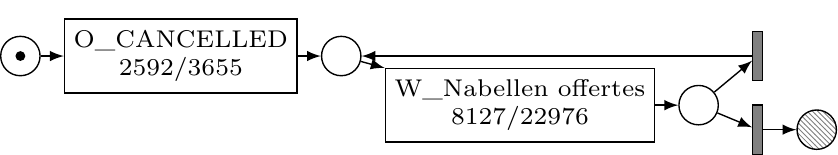}
	}
	
	\subfloat[\label{sfig:bpic-lpm_4}support: 4749]{
		\includegraphics{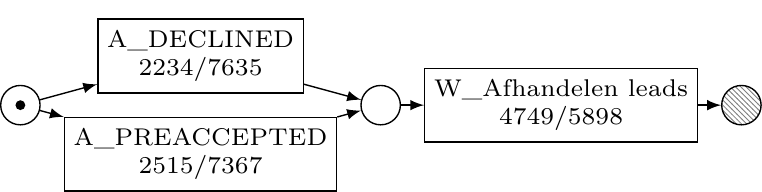}
	}
	
	\subfloat[\label{sfig:bpic-lpm_5}support: 606]{
		\includegraphics{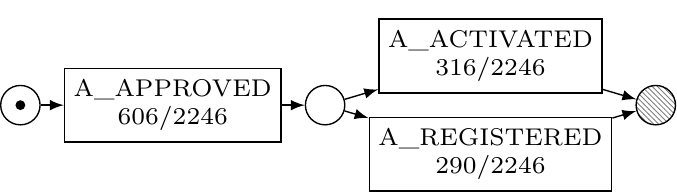}
	}
	
	\subfloat[\label{sfig:bpic-lpm_6}support: 13087]{
		\includegraphics{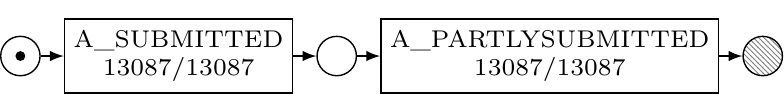}
	}
	\caption{The LPM set that is mined with the greedy F-score based selection approach from the BPI'12 dataset.}
	\label{fig:bpi12-results}
	
\end{figure}

Figure~\ref{fig:bpi12-results} shows the obtained LPM set, which consists of six patterns. The LPM set has a coverage of 0.447, indicating that almost half of the events in the sequence database are described by one of the six patterns. The precision of the LPM set is 0.359, resulting in an F-score of 0.3983. 

The first LPM (in Figure~\ref{fig:bpi12-results}a) shows that \emph{O\_SELECTED} and \emph{A\_FINALIZED} are executed concurrently and are ultimately followed by \emph{O\_CREATED} and \emph{O\_SENT}. This pattern occurs 5015 times in the sequence database, which corresponds to occurrences of \emph{O\_SELECTED} adhering to this behavior. For the other activities, \emph{A\_FINALIZED}, \emph{O\_CREATED}, and \emph{O\_SENT} a large majority of 5015 out of the 7030 events of each of these activities are described by this behavior. 

Figure~\ref{fig:bpi12-results}b shows that \emph{O\_ACCEPTED} is generally followed by both \emph{A\_REGISTERED} and \emph{A\_ACTIVATED}, but the order in which these two activities occur is not consistent. Finally, this behavior is followed by one or more occurrences of activity \emph{W\_Valideren aanvraag} (in English: ``Validate request''). 

Figure~\ref{fig:bpi12-results}c shows a pattern where \emph{O\_CANCELLED} is generally followed by multiple occurrences of \emph{W\_Nabellen offertes} (in English: ``Calling after offers''). Looking at the numbers (2592 \emph{O\_CANCELLED} and 8127 \emph{W\_Nabellen offertes} events), we can conclude that on average the call center employees can the customer three times after an offer has been canceled.

Figure~\ref{fig:bpi12-results}d shows that \emph{W\_Afhandelen leads} (in English: ``Processing leads''), are in 4749 out of 5898 cases preceded either by an \emph{A\_DECLINED} (2234 times) or by \emph{A\_PREACCEPTED} (2515 times).

Figure~\ref{fig:bpi12-results}e shows that \emph{A\_APPROVED} is sometimes (606 out if 2246 times) followed by \emph{A\_ACTIVATED} (316 times) or by \emph{A\_REGISTERED} (290 times). 

Finally, Figure~\ref{fig:bpi12-results}f shows that all occurrences of \emph{A\_SUBMITTED} are followed by \emph{A\_PARTLYSUBMITTED}, and that, likewise, all occurrences of \emph{A\_PARTLYSUBMITTED} are preceded by \emph{A\_SUBMITTED}.

\section{Conclusion}
\label{sec:conclusions}
The foremost contribution of this paper is a set of techniques to mine a non-redundant set of generalizing patterns captured as \emph{Local Process Models} (LPMs) from a sequence database. We have shown that the generalizing capabilities of LPMs, coupled with the non-redundancy property, allow us to concisely summarize the sequence database via a small number of patterns compared to alternative methods. The paper also introduced a framework for evaluating the quality of sets of LPMs, which takes into account both the share of the events in the sequence database that are \emph{covered} by an LPM set as well as \emph{the degree of redundancy} in the LPM set.

We have outlined multiple approaches to the posed problem, which differ in computational complexity and in their trade-off between \emph{how many events are covered by the pattern set} and emph{the degree of redundancy of the pattern set}. We have shown that alignment-based selection can reduce the redundancy of a set of LPMs without loss in coverage. The greedy selection and the greedy F-score-based selection approach are able to achieve further reduction of redundancy at the cost of coverage.

The proposed method for mining LPM patterns has manifold applications. A direct application that we envisage is to discover repetitive routines from fine-grained event (e.g. clickstreams), which may be amenable to automation. Other applications including exploratory analysis of smart home data in order to extract daily routines, as well as analysis of Web usage logs in order to identify typical usage patterns which may give rise to opportunities for Web site optimization. Investigating these applications is an avenue for future work.

\bibliographystyle{spbasic}      
\bibliography{paper}   

\end{document}